\documentclass[%
 reprint,
 superscriptaddress,
 longbibliography,
 preprintnumbers,
nobibnotes,
 amsmath,amssymb,
aps,
 prl,
floatfix,
]{revtex4-2}
\usepackage[utf8]{inputenc}
\usepackage{comment}
\usepackage[T1]{fontenc}
\usepackage{graphicx}
\usepackage{dcolumn}
\usepackage{bm}
\usepackage{hyperref}
\usepackage{xcolor}
\usepackage{braket}
\usepackage{mathtools}
\usepackage{subfigure}
\usepackage[version=4]{mhchem}

\usepackage{physics}
\usepackage{siunitx}
\usepackage[mode=buildnew]{standalone}

\usepackage[maxfloats=256]{morefloats}
\definecolor{mediumorchid}{RGB}{186, 85, 211}
\hypersetup{
    colorlinks=true,
    linkcolor=mediumorchid,
    citecolor=mediumorchid,
    filecolor=mediumorchid,      
    urlcolor=mediumorchid,
    }

\begin{document}

\preprint{}

\title{Cavity QED Control of Quantum Hall Stripes}
\date{\today}

\author{Lorenzo Graziotto}
\email{lgraziotto@phys.ethz.ch}
\affiliation{Institute for Quantum Electronics, ETH Zürich, Zürich 8093, Switzerland}
\affiliation{Quantum Center, ETH Zürich, Zürich 8093, Switzerland}
\author{Josefine Enkner}
\affiliation{Institute for Quantum Electronics, ETH Zürich, Zürich 8093, Switzerland}
\affiliation{Quantum Center, ETH Zürich, Zürich 8093, Switzerland}
\author{Sambuddha Chattopadhyay}
\affiliation{Institute for Theoretical Physics, ETH Zürich, Zürich 8093, Switzerland}
\affiliation{Lyman Laboratory, Department of Physics, Harvard University, Cambridge MA, USA}
\author{Jonathan B. Curtis}
\affiliation{Institute for Theoretical Physics, ETH Zürich, Zürich 8093, Switzerland}
\author{Ethan Koskas}
\affiliation{Institute for Quantum Electronics, ETH Zürich, Zürich 8093, Switzerland}
\affiliation{Quantum Center, ETH Zürich, Zürich 8093, Switzerland}
\author{Christian Reichl}
\affiliation{Laboratory for Solid State Physics, ETH Zürich, Zürich 8093, Switzerland}
\affiliation{Quantum Center, ETH Zürich, Zürich 8093, Switzerland}
\author{Werner Wegscheider}
\affiliation{Laboratory for Solid State Physics, ETH Zürich, Zürich 8093, Switzerland}
\affiliation{Quantum Center, ETH Zürich, Zürich 8093, Switzerland}
\author{Giacomo Scalari}
\affiliation{Institute for Quantum Electronics, ETH Zürich, Zürich 8093, Switzerland}
\affiliation{Quantum Center, ETH Zürich, Zürich 8093, Switzerland}
\author{Eugene Demler}
\affiliation{Institute for Theoretical Physics, ETH Zürich, Zürich 8093, Switzerland}
\author{Jérôme Faist}
\email{jerome.faist@phys.ethz.ch}
\affiliation{Institute for Quantum Electronics, ETH Zürich, Zürich 8093, Switzerland}
\affiliation{Quantum Center, ETH Zürich, Zürich 8093, Switzerland}

\begin{abstract}
    Controlling quantum phases of materials with vacuum field fluctuations in engineered cavities is a novel route towards the optical control of emergent phenomena. We demonstrate, using magnetotransport measurements of a high-mobility two-dimensional electron gas, striking cavity-induced anisotropies in the electronic transport, including the suppression of the longitudinal resistance well below the resistivity at zero magnetic field. Our cavity-induced effects occur at ultra-low temperatures ($< \SI{200}{mK}$) when the magnetic field lies between quantized Hall plateaus. We interpret our results as arising from the stabilization of thermally-disordered quantum Hall stripes. Our work presents a clear demonstration of the cavity QED control of a correlated electronic phase.
\end{abstract}

\maketitle

The prospect of optically inducing correlated electronic phases of matter on-demand in solid-state systems has materialized in the past decade due to rapid developments in the tailoring of electronic properties with strong electromagnetic fields~\cite{basov2017towards}. A nascent complementary program is the control of quantum materials using the vacuum fields of engineered cavities, a mode of passive control in which striking effects can be achieved while maintaining the system \textit{at equilibrium}~\cite{garcia2021manipulating, hubener2021engineering, lu2025cavity}. Central to this approach is the notion that empty space contains vacuum fluctuations \cite{milonni2013quantum} which give rise to canonical quantum electrodynamics effects such as the Casimir force~\cite{casimir1948attraction} or the Lamb shift~\cite{lambShift}. Shaping the electromagnetic environment by designing suitable resonators thus allows one to harness vacuum fields to influence material properties, an idea that has been explored theoretically in diverse contexts ranging from ferroelectricity~\cite{ashida2020quantum, curtis2023local,Latini.2021} to superconductivity~\cite{schlawin2019cavity,Sentef.2018,Curtis.2019,Allocca.2019,Raines.2020} to ferromagnetism~\cite{roman2021photon}. Cavity effects, leveraging either thermal or vacuum fields, have recently been experimentally demonstrated by controlling the metal-to-insulator transition temperature in 1T-\ce{TaS1}~\cite{jarc2023cavity}, and by altering the transport properties of the integer and fractional quantum Hall effect in high-mobility two-dimensional electron systems (2DESs) at millikelvin temperatures~\cite{appugliese2022breakdown, enkner2024enhanced}.

The quantum Hall system, which is realized by subjecting a 2DES to a perpendicular magnetic field $B$~\cite{giuliani2005quantum}, represents an ideal playground for vacuum cavity control as the effective light-matter interaction length scales---set by the radius of the cyclotron orbit---are larger than in a typical solid by four orders of magnitude~\cite{Girvin_Yang_2019}. Additionally, the magnetic field quenches the kinetic energy of the system, leading to a number of energetically competing, correlated electronic phases defined by strong Coulomb interactions that are amenable to cavity control. Crucially, the correlated phases exhibited by the quantum Hall system depend on the number of occupied Landau levels (LLs)---which constitute its distinctive energy spectrum---quantified by the filling factor $\nu = h n_s / e B$, where $h$ is Planck’s constant, $-e$ the electron charge, and $n_s$ indicates the two-dimensional (2D) electron density. Near $\nu=N+1/2$, with $N>4$ an integer, one such correlated electronic phase, known as quantum Hall stripes, arises. 

Quantum Hall stripes are a form of electronically-driven charge density wave order which manifests at ultralow temperatures, in half-filled high LLs. Electronic density modulation at the scale of the cyclotron radius ($\sim \SI{50}{nm}$ at $\nu = 8+1/2$)---arising from the ring-like shape of the higher-Landau-level wave functions~\cite{fogler2002stripe, fradkin2010nematic}---becomes thermodynamically favorable below $\sim \SI{1}{K}$. However, while energetic constraints select the wavelength of the density modulation, they do not discriminate between different orientations of the stripe order. Thus, in homogeneous and isotropic electron gases---such as those used in our study---thermal fluctuations scramble the orientation of the stripes, precluding their direct observation in magnetotransport measurements. Experimental signatures of stripes are observed when structural anisotropies aligned with a particular crystallographic direction of the heterostructure~\cite{cooper2001investigation}, induced by strain~\cite{koduvayur2011effect}, or in-plane magnetic fields~\cite{pan2000reorientation, shi2017effect} result in the alignment of the stripes on a macroscopic scale, giving rise to striking magnetotransport anisotropies measured near high half-integer filling factors~\cite{lilly1999evidence, du1999strongly}.

\begin{figure*}[!hbt]
    \centering
    \includegraphics[width=\textwidth]{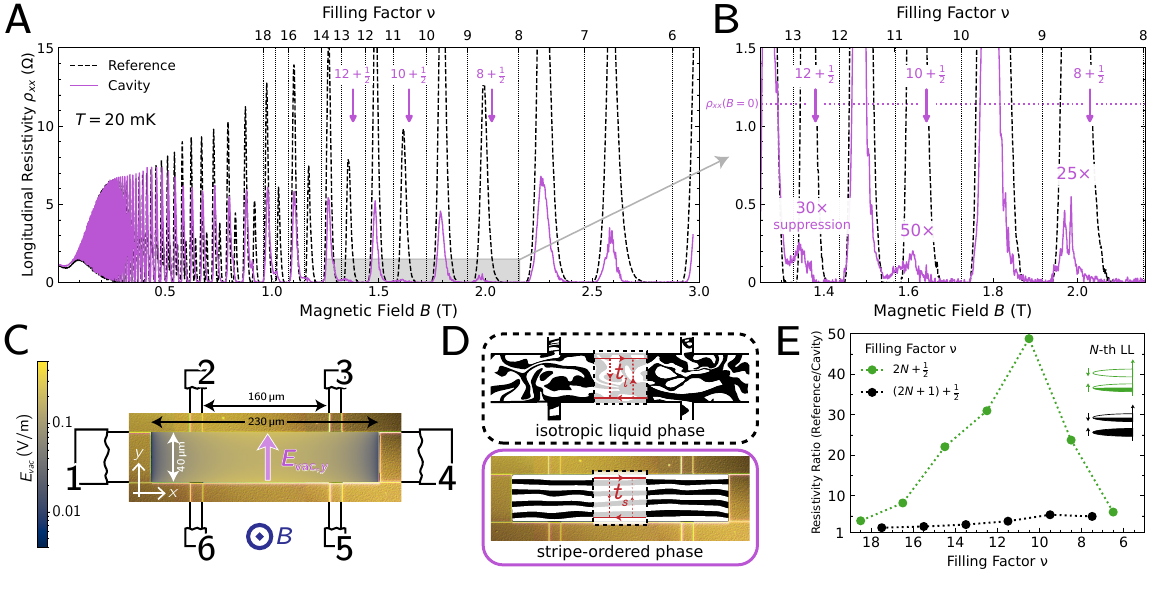}
    \caption{\textbf{Cavity-induced stripe-ordered phase in the vacuum field of a slot antenna} \\ (\textbf{A}) Longitudinal resistivity $\rho_{xx}$ as a function of magnetic field $B$ for a cavity-embedded (purple solid line) and a reference Hall bar (HB) (black dashed line), measured along the $\hat{\vb{x}}$ direction---according to the axes shown in \textbf{C}. The top axis reports the filling factor $\nu$ corresponding to $B$ at the 2DES density \SI{4e11}{cm^{-2}}. (\textbf{B}) Enlargement of the cavity-induced suppressed resistivity at filling factors $12+1/2$, $10+1/2$, and $8+1/2$ (grey rectangle in \textbf{A}), showing $30\times$, $50\times$, and $25\times$ reductions, respectively, as compared to the resistivity measured in the reference HB.
    (\textbf{C}) Optical microscope picture of the slot antenna resonator, which consists of a $\SI{230}{\micro m}\times\SI{40}{\micro m}$ rectangular cutout from a metal plane (gold color) evaporated on top of the HB. The gradient color represents the vacuum electric field polarized in the $\hat{\vb{y}}$ direction of the cavity fundamental mode, as obtained from finite element simulations. The contact leads and continuation of the \SI{40}{\micro m}-wide HB are visible as glowing lines below the metal.
    (\textbf{D}) Cartoon of the two different phases of the 2DES without (top) and within (bottom) the cavity. In the middle of each HB the basic process which scatters edge states traveling in opposite directions into each other is sketched: in the stripe-ordered phase the backscattering amplitude $1-t_s$ is strongly reduced. (\textbf{E}) Ratio between the resistivity $\rho_{xx}$ measured in the reference and in the cavity-embedded HBs, as a function of filling factor. We observe a distinct behavior when only the energetically lower spin-resolved Landau level (LL) is partially filled (green markers) or when the lower is completely filled and the upper is partially filled (black markers). The inset shows a sketch of the electronic occupation of the $N$-th LL.}
    \label{fig:panel1}
\end{figure*}

Here we demonstrate the cavity QED control of quantum Hall stripes by means of the vacuum electromagnetic field of a slot antenna cavity, which was designed and engineered to realize a strongly anisotropic coupling with the 2DES, capable of steering the stripe order. Using our cavity, we induce not only striking anisotropies in the longitudinal transport, but also the nearly complete zeroing of the longitudinal resistivity {\em far below its value in the absence of magnetic field}, away from quantizing magnetic field values. No other anisotropy-inducing mechanism~\cite{lilly1999evidence, du1999strongly, pan2000reorientation, cooper2001investigation, koduvayur2011effect, shi2017effect} has ever been demonstrated to achieve such a dramatic effect, which provides strong evidence of the capability of spatially-structured cavity vacuum fluctuations to \emph{improve} the transport in a correlated electronic phase. 

In our experiment, we employ a 2DES with mobility $\mu = \SI{2e7}{cm^2 V^{-1} s^{-1}}$ and density $n_s = \SI{4e11}{cm^{-2}}$, realized in a high-quality epitaxially grown GaAs-based heterostructure, featuring excellent homogeneity and isotropy as discussed in the Supplementary Material (SM)~\cite{supplementary}. We fabricate a \SI{40}{\micro m}-wide Hall bar (HB) embedded in the slot antenna cavity, and we measure its magnetotransport properties at high half-integer filling factors and ultralow temperatures, where vacuum fluctuations of the cavity field vastly dominate thermal fluctuations, as the photon population is below~$10^{-6}$ at the relevant temperatures. In Fig.~\ref{fig:panel1}A,B, we report the main observation of our work: the striking, 50-fold suppression of the longitudinal resistivity, measured along the $\hat{\vb{x}}$ direction at filling factor $\nu = 10+1/2$ and temperature \SI{20}{mK}, in the cavity-embedded HB, as compared to the reference HB. The latter was fabricated on the same chip---physically separated by a distance of about \SI{2.5}{mm}---and measured in the same cool down. The fact that the longitudinal resistivity is suppressed down to~\SI{0.2}{\ohm}, well below its value at zero field of~\SI{1.15}{\ohm}, demonstrates a suppression of backscattering away from quantized magnetic field values, and points to the stabilization by the cavity of a continuous stripe along the whole \SI{160}{\micro m}-distance between the voltage probes (2-3 in Fig.~\ref{fig:panel1}C).

\begin{figure*}[hbt]
    \centering
    \includegraphics[width=\textwidth]{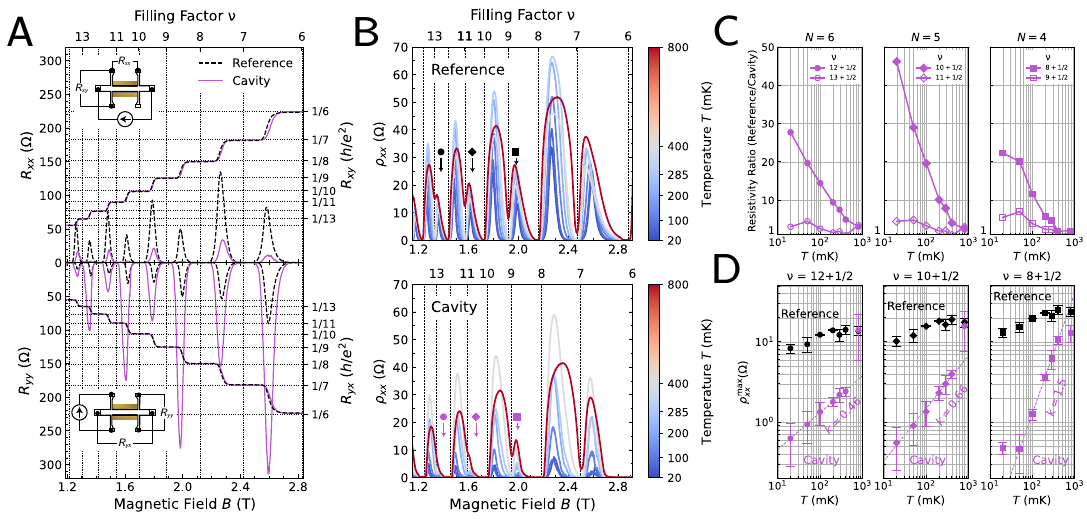}
    \caption{\textbf{Magnetotransport in the stripe-ordered phase} \\ (\textbf{A}) Longitudinal and transverse resistances measured on a reference (black dashed lines) and on a cavity-embedded HB (purple solid lines), as a function of perpendicular magnetic field, at \SI{20}{mK} temperature. The left axes refer to the longitudinal resistances $R_{xx}$ (top) and $R_{yy}$ (bottom, with inverted axis), while the right axes refer to the transverse resistances $R_{xy}$ (top) and $R_{yx}$ (bottom, with inverted axis). The insets show the measurement schemes employed to measure the different resistances. (\textbf{B}) Longitudinal resistivity $\rho_{xx} = R_{xx}/4$ as a function of magnetic field for the reference (top) and cavity-embedded HB (bottom), measured at different temperatures of the mixing-chamber plate (colors according to the color bar on the right). (\textbf{C}) Ratio between the resistivity $\rho_{xx}$ measured in the reference and in the cavity-embedded HBs as a function of temperature, for filling factors $2N+1/2$ (full markers) and $(2N+1)+1/2$ (empty markers). The data are taken from the plots in \textbf{B}: circle, diamond, and square markers refer to $N=6,5,4$, respectively, as indicated on top of the three subplots. (\textbf{D}) Longitudinal resistivity maxima at half-integer filling factors as a function of temperature, in log-log scale. Again, the data are taken from the plots in \textbf{B}: circles, diamonds, and squares refer to filling factor $12+1/2,10+1/2$, and $8+1/2$, respectively, while black and purple colors refer to the reference and cavity sample, respectively. The resistivity maxima of the cavity sample follow a power-law behavior as a function of temperature, $\rho_{xx}^\mathrm{max} \propto T^k$, with $k$ indicated in the subplots.}
    \label{fig:panel2}
\end{figure*}

The substantial subwavelength confinement of the ground-state electromagnetic modes realized in the slot antenna cavity is a central ingredient in magnifying the amplitude of the vacuum fluctuations and hence their coupling to the 2DES, which is usually quantified by the normalized light-matter coupling $\eta$. It was shown in Ref.~\cite{paravicini2019magneto} that in this system the effective cavity volume is $7\times10^{-4}(\lambda_{\SI{205}{GHz}}/2)^3$ for the fundamental mode with frequency \SI{205}{GHz}~\cite{paravicini2019magneto}, and that $\eta \sim 20\%$: the light-matter coupled system is thus said to be in the \emph{ultrastrong} coupling regime~\cite{frisk2019ultrastrong}. The geometry of the cavity was chosen such to provide a large number of modes (five) between \qtyrange{0.2}{1.2}{THz} having electric field polarized along the $\hat{\vb{y}}$ direction (see Fig.~\ref{fig:panel1}C), perpendicular to the edges of the resonator. As we show theoretically below, this favors the alignment of the stripes parallel to the edges---i.e.\ with the wavevector of the charge density modulation parallel to the field polarization---rendering the transport ``easier'' along the $\hat{\vb{x}}$ direction, and ``harder'' along $\hat{\vb{y}}$. A cartoon of the physical mechanism at play is depicted in Figure~\ref{fig:panel1}D, contrasting the stripe-ordered phase of the 2DES inside the cavity with the isotropic liquid phase manifested in the absence of it. 

To realize and maintain the stripe order over the whole length of \SI{160}{\micro m} which separates the voltage probes, we have crafted our slot antenna endowing it with smooth cutout edges at a submicrometer scale. Without this care, the dramatic reduction below the zero-field value is unattainable (a similar but lower suppression is observed also in 2DESs having different densities hosted in different heterostructures, and in our hovering cavity experiment~\cite{enkner2024enhanced}---see SM~\cite{supplementary}). To confirm this statement, we also fabricated and investigated cavities having stepped or serrated edges by design~\cite{supplementary}. We notice that the reduced resistivity in the cavity sample is not merely due to sample-specific features such as the different disorder configurations in the cavity-embedded and reference HBs, as the zero-field resistivity is not modified by the presence of the cavity. The observed suppression is robust: it is measured in both magnetic field directions, and for increasing and decreasing field sweeps~\cite{supplementary}. 

As shown in Fig.~\ref{fig:panel1}E, we observe the resistivity suppression, albeit to a lesser extent, at all half-integer filling factors between $6+1/2$ and $18+1/2$, with a consistent distinction between $\nu=2N+1/2$, with $N\geq3$ an integer, and $\nu=(2N+1)+1/2$, where it amounts to about an order of magnitude less. We connect the latter observation to a different occupation of the spin-resolved $N$-th LL: while at $\nu=2N+1/2$ only one spin polarization is present, at $\nu=(2N+1)+1/2$ the energetically lower spin-resolved level is full, and electrons with the opposite spin partially fill the upper level. An analogous spin dependence was observed also in Refs.~\cite{lilly1999evidence, du1999strongly}, and in Ref.~\cite{wexler2001disclination} a lower critical temperature of the stripe-ordered phase was computed for the $\nu=(2N+1)+1/2$ case as opposed to $\nu=2N+1/2$. Since the critical temperature is linked to the exchange energy magnitude, we also mention that a strong reduction of the effective $g$-factor due to cavity vacuum fields was reported in Ref.~\cite{enkner2024enhanced} and also observed in our sample~\cite{supplementary}.

We characterize the anisotropic magnetotransport in the stripe-ordered phase by comparing the longitudinal resistance $R_{xx}$, measured along the $\hat{\vb{x}}$ direction (Fig.~\ref{fig:panel2}A, top panel), with the resistance $R_{yy}$, measured via the scheme depicted in the inset, which quantifies the transport in the orthogonal $\hat{\vb{y}}$ direction (bottom panel). In complete agreement with our interpretation, we observe an \emph{increase} in $R_{yy}$ of more than a factor of 5 in the cavity-embedded HB at the same filling factors $8+1/2, 10+1/2$ for which the longitudinal resistance $R_{xx}$ is \emph{suppressed}---as compared to the reference HB. As discussed in detail in the SM~\cite{supplementary}, the increase by a factor of 5 instead of 50 is due to the so-called nonlocal edge-states contribution to $R_{yy}$~\cite{wang1992measurements}. As expected, the transverse resistance displays well quantized plateaus~\cite{enkner2024testing}. The cavity and reference differ slightly only in the plateau-to-plateau transition, due to the minimal density difference of the two HBs---estimated to be about 1\% by fitting the $R_{xy}$ slope near zero field. 

We investigate the appearance of the stripe order through the temperature dependence of the longitudinal resistivity. As shown in Fig.~\ref{fig:panel2}B,C, we observe the cavity-induced transport signatures only at ultralow temperatures: the resistivity at \SI{800}{mK} is the same between the cavity-embedded HB and the reference. We further notice the striking difference in the temperature behavior of the resistivity maxima at half-integer filling factors in the reference and cavity sample (Fig.~\ref{fig:panel2}D): while the former shows at most a twofold increase with increasing temperature, the latter climbs as a power-law over an order of magnitude between 20 and \SI{500}{mK}.


We remark in passing that in the present work we focus on correlated, low-temperature transport at high \textit{half-integer} filling factors, where the 2DES behaves as an isotropic liquid in the absence of a cavity (Fig.~\ref{fig:panel1}B), differently from our previous works reported in Refs.~\cite{appugliese2022breakdown, enkner2024enhanced}. 

\begin{figure*}[!hbt]
    \centering
    \includegraphics[width=\textwidth]{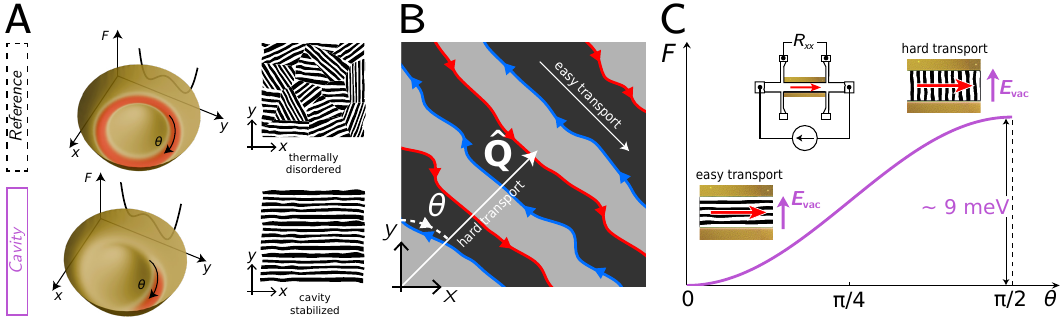}
    \caption{\textbf{Orientational stabilization of fluctuating stripe order by the cavity} \\ (\textbf{A}) Free energy surface $F(\theta)$ as a function of the modulation direction of the stripe-ordered phase (displaced from the origin of the axes for visual clarity). Notice that a stripe modulation at angle $\theta$ means that the stripes are aligned along $\theta-\SI{90}{\degree}$. In the reference case (top) the surface is rotationally invariant, so that stripes order is formed with arbitrary orientation $\theta$, and thermally fluctuates (red shading of the $F$ surface): domains of stripes with different orientations form in the reference sample, and no macroscopic orientation is present. In the presence of the cavity (bottom) the $F$ surface possesses a clear minimum at $\theta=0$, which macroscopically aligns the stripes. (\textbf{B}) The macroscopic stripe alignment, with modulation wavevector $\hat{\vb{Q}}$, defines an axis along which transport is ``easy''---orthogonal to $\hat{\vb{Q}}$---and one along which transport is ``hard''---parallel to $\hat{\vb{Q}}$. (\textbf{C}) Fixing the direction of the polarized vacuum electric field ($\theta=0$), the lowest free energy is obtained with a modulation parallel to it, hence resulting in easy transport in the $\hat{\vb{x}}$ direction, which is what is measured in our experiment.}
    \label{fig:panel3}
\end{figure*}



We interpret our measurements by proposing that the anisotropic vacuum fluctuations of the electric field of the slot antenna cavity fundamental mode energetically favor the alignment of preexisting but orientationally disordered local stripe order along the $\hat{\vb{x}}$-axis (i.e.\ charge order with density modulation with wavevector $\hat{\vb{Q}}$ pointing along $\hat{\vb{y}}$). Such an interpretation provides a consistent picture for the main qualitative findings of our experiment. First, if stripes form along the $\hat{\vb{x}}$-axis, then transport is ``easy'' along this direction, explaining the dramatic cavity suppression of the longitudinal resistivity~$\rho_{xx}$. Second, stripes along the $\hat{\vb{x}}$-axis imply that transport along the $\hat{\vb{y}}$-axis is ``hard'' as it requires disorder-induced scattering processes across the stripes, justifying the increase in $R_{yy}$. Third, stripes give rise to anisotropic magnetotransport, as observed in our experiment, in the vicinity of high half-integer filling factors (e.g.\ $\nu = 8+1/2, 10+1/2$). Fourth, the appearance of the cavity-induced signatures at ultralow temperatures \qtyrange{20}{100}{mK}, coupled with their disappearance near \SI{1}{K}, match the temperature scales at which stripes appear and disappear, respectively, in the samples in which they have been observed~\cite{lilly1999evidence, du1999strongly}. Finally, we underscore that magnetotransport in the sample is not intrinsically anisotropic, as discussed in the SM~\cite{supplementary}: the presumptive stripe order is manifestly a cavity-induced effect.

To conceptualize why the electromagnetic vacuum fluctuations of the cavity orient stripes along the $\hat{\vb{x}}$-axis, we estimate the anisotropy in free energy arising from the interaction between a fixed orientation of stripes and the vacuum fluctuations of the electric field of the cavity. Physically, this can be interpreted as the change in the Casimir energy~\cite{bordag2009advances} of the slot antenna cavity due to a particular orientation of electronic stripe order which anisotropically modifies the refractive index of the system. Within the Matsubara formalism~\cite{altland2023condensed}, we find that the free energy of a specific orientation of stripes is given by:
\begin{equation}
\label{eqn:cavFE}
F(\theta) = \frac{1}{\pi} \sum_{a} \int_{0}^{\infty}  d\omega  \, \omega \sigma_{aa
}(i \omega)(\theta)  \overline{\langle \textbf{A}^{a}(i \omega) \textbf{A}^{a}(-i \omega) \rangle}, 
\end{equation}
where the dynamical conductivity tensor---continued to imaginary frequencies---of stripes with modulation wavevector at an angle $\theta$ with respect to the cavity, is given by $\sigma_{ij}(i \omega)(\theta)$ and where the Matsubara frequency correlation function of the vector potential arising from the slot antenna is given by  $\langle \textbf{A}^a(i\omega) \textbf{A}^b(-i\omega) \rangle = \int d^2 \textbf{R} \langle \textbf{A}^a(\mathbf{R},i\omega) \textbf{A}^b(\mathbf{R},-i\omega) \rangle$ (see SM for details~\cite{supplementary}). 

Our formula suggests that the stripe configuration with the lowest free energy is the one for which the hard axis of the conductivity---the axis for which the conductivity (resistivity) is smaller (larger)---is aligned with the axis of the resonator which experiences the largest amount of vacuum electromagnetic fluctuations. As the ultrastrongly coupled fundamental mode is polarized along $\hat{\vb{y}}$, the cavity forces stripes to have their hard axis along the $\hat{\vb{y}}$ direction, as consistent with our experimental observation. Within a canonical model of transport in the stripe phase~\cite{MacDonald.2000}, we estimate that the anisotropy favoring alignment along the $\hat{\vb{x}}$-axis amounts to a collective free energy difference between orthogonal ($\theta = 0$ and $\theta = \pi/2$) orientations of \SI{9}{meV}, for the $\sim 10^{6}$ electrons that constitute the highest partially filled Landau level, in qualitative agreement with previous estimates~\cite{ivar} of the anisotropy required to reorient stripes (see Fig.~\ref{fig:panel3}C). 

Cavity QED \textit{control} entails the intentional design of the electromagnetic environment to manipulate the emergent electronic properties of a quantum material. In this work, we exploited the anisotropic vacuum fluctuations of a terahertz slot antenna cavity to stabilize quantum Hall stripes in a specific orientation, and we showcased the resultant striking suppression of longitudinal resistivity and magnetotransport anisotropies. Together with our functional understanding of the role of the cavity in structuring correlated order, we advance that this work represents a clear demonstration of the cavity QED control of a quantum material. As has proven successful in the broader optical control program~\cite{disa2021engineering}, cavities may be used to selectively stabilize fluctuating order in electronic systems. Our demonstration suggests to apply this approach to mesoscopic systems---in particular moiré materials~\cite{Moire}---which harbor many of the same characteristics that make quantum Hall a model setting for cavity control: crowded phase diagrams arising from the elevated importance of electronic interactions in the presence of quenched kinetic energy and large effective dipole sizes. While magnetotransport anisotropy has predominantly been used to diagnose quantum Hall stripes, complementary approaches could be used. Further evidence could be generated by examining the collective modes of the 2DES, exploring their plausible anisotropic dispersion. Previously, the collective mode spectrum of stripes has been mapped out using a combination of microwave driving with surface acoustic waves~\cite{MicrowaveSAW}, although scanning near-field optical microscopy (SNOM) techniques could also be applied~\cite{liu2016nanoscale}. Direct confirmation may come from real-space imaging: single electron transistors (SET) measurements~\cite{ilani2004microscopic} have accessed the relevant \SI{100}{nm}-length scales at which stripes are expected to form.

\section*{Acknowledgements}
We thank Gian Lorenzo Paravicini-Bagliani and Felice Appugliese for processing and measuring previous samples~\cite{supplementary}, where a similar magnetotransport anisotropy was first observed. We thank Peter Märki for technical support. We thank Junkai Dong and Frieder Lindel for discussions. This work was funded by the Swiss National Science Foundation (SNF) (Grant number 10000397).

\bibliography{bibliography,refs-curtis}
\end{document}


\onecolumngrid

\preprint{}

\title{\textsc{Supplementary Material} \\[0.5cm] Cavity QED Control of Quantum Hall Stripes}

\author{Lorenzo Graziotto}
\email{lgraziotto@phys.ethz.ch}
\affiliation{Institute for Quantum Electronics, ETH Zürich, Zürich 8093, Switzerland}
\affiliation{Quantum Center, ETH Zürich, Zürich 8093, Switzerland}
\author{Josefine Enkner}
\affiliation{Institute for Quantum Electronics, ETH Zürich, Zürich 8093, Switzerland}
\affiliation{Quantum Center, ETH Zürich, Zürich 8093, Switzerland}
\author{Sambuddha Chattopadhyay}
\affiliation{Institute for Theoretical Physics, ETH Zürich, Zürich 8093, Switzerland}
\affiliation{Lyman Laboratory, Department of Physics, Harvard University, Cambridge MA, USA\looseness=-1}
\author{Jonathan B. Curtis}
\affiliation{Institute for Theoretical Physics, ETH Zürich, Zürich 8093, Switzerland}
\author{Ethan Koskas}
\affiliation{Institute for Quantum Electronics, ETH Zürich, Zürich 8093, Switzerland}
\affiliation{Quantum Center, ETH Zürich, Zürich 8093, Switzerland}
\author{Christian Reichl}
\affiliation{Laboratory for Solid State Physics, ETH Zürich, Zürich 8093, Switzerland}
\affiliation{Quantum Center, ETH Zürich, Zürich 8093, Switzerland}
\author{Werner Wegscheider}
\affiliation{Laboratory for Solid State Physics, ETH Zürich, Zürich 8093, Switzerland}
\affiliation{Quantum Center, ETH Zürich, Zürich 8093, Switzerland}
\author{Giacomo Scalari}
\affiliation{Institute for Quantum Electronics, ETH Zürich, Zürich 8093, Switzerland}
\affiliation{Quantum Center, ETH Zürich, Zürich 8093, Switzerland}
\author{Eugene Demler}
\affiliation{Institute for Theoretical Physics, ETH Zürich, Zürich 8093, Switzerland}
\author{Jérôme Faist}
\email{jerome.faist@phys.ethz.ch}
\affiliation{Institute for Quantum Electronics, ETH Zürich, Zürich 8093, Switzerland}
\affiliation{Quantum Center, ETH Zürich, Zürich 8093, Switzerland}

\begin{abstract}
    
\end{abstract}

\maketitle
\tableofcontents


\section{Material and methods}

\subsection{Heterostructure details}
The two-dimensional electron system (2DES) is hosted in a \SI{24}{nm}-wide GaAs/AlGasAs quantum well grown via molecular-beam epitaxy (MBE) in the Laboratory for Solid State Physics at ETH Zürich (the heterostructure is labelled D170202B, and we internally refer to the sample as \textbf{D170202B-2}). The well is modulation doped, with a \SI{70}{nm} spacer between the doping layer and the edge of the barrier, and overall the 2DES is located \SI{130}{nm} below the surface of the heterostructure. The carrier density and mobility measured at \SI{1.3}{K} without illumination are $n_s = \SI{3.98e11}{cm^{-2}}$, $\mu = \SI{2.03e7}{cm^2 V^{-1} s^{-1}}$, respectively. At \SI{}{mK} temperatures from the Hall resistance slope we obtain a density varying between \SI{4.17e11}{} and \SI{4.36e11}{cm^{-2}} across the sample, increasing over a length scale of about \SI{3.3}{mm}. The isotropy of the bulk resistivity is discussed in Section~\ref{sec:squarepatch}, and we discuss in Section~\ref{sec:density_inhom} the procedure employed to account for density gradients in the estimate of the longitudinal resistivities from the Hall bar (HB) resistance measurements. 

\subsection{Sample fabrication}
The samples are processed via standard photolithography techniques in a clean-room environment. The procedure is carried out on a $9\times10\,\SI{}{mm^2}$ chip cleaved from the MBE-grown wafer, and it consists of the following steps:
\begin{enumerate}
    \item Definition of the Hall bars and of the square patch using positive resist (AZ1505) and high-resolution direct laser writing;
    \item Etching of the mesa region using a diluted piranha solution (\ce{H2SO4}:\ce{H2O2}(30\%):\ce{H2O} 1:8:60);
    \item Definition of the contacts using negative image-reversal resist (AZ5214E) and high-resolution direct laser writing;
    \item Evaporation of \ce{Ge}/\ce{Au}/\ce{Ge}/\ce{Au}/\ce{Ni}/\ce{Au} (26/53/26/53/40/50 \SI{}{nm}) eutectic mixture, lift-off and annealing (\SI{500}{\celsius} for \SI{300}{s});
    \item Definition of the resonator plane using negative image-reversal resist (AZ5214E) and high-resolution direct laser writing;
    \item Evaporation of \ce{Ti}/\ce{Au} (10/200 \SI{}{nm}) and lift-off.
\end{enumerate}

\subsection{Measurement scheme}
The transport measurements are performed with state-of-the-art techniques optimized for investigating the quantum Hall effect in two-dimensional electron systems~\cite{baer2015transport}. We cool the samples down to \SI{12}{mK} with a Bluefors dilution refrigerator. The voltage is measured with commercial MFLI Zurich Instruments digital lock-in amplifiers, in conjunction with custom-made differential ac low-noise preamplifiers, which increase the signal by a factor of $1000$. We inject current symmetrically by applying an ac-modulated voltage of \SI{2}{V} root-mean-square (rms) at the demodulation frequency \SI{13.333}{Hz} to two \SI{100}{M\ohm} in front of the sample's contacts (or equivalently applying \SI{0.2}{V} to two \SI{10}{M\ohm}), such that a \SI{10}{nA} rms current circulates in the circuit, a low value chosen to limit electronic heating of the sample. To demodulate the lock-in input signal, we emply a fourth-order low-pass filter with time constant of \SI{0.1}{s}. Before the contacts to the sample, low-pass \SI{100}{kHz} filters are placed with the purpose of minimizing electric spikes or heating effects from the measurement setup. 

\begin{figure}
    \centering
    \includegraphics[width=\linewidth]{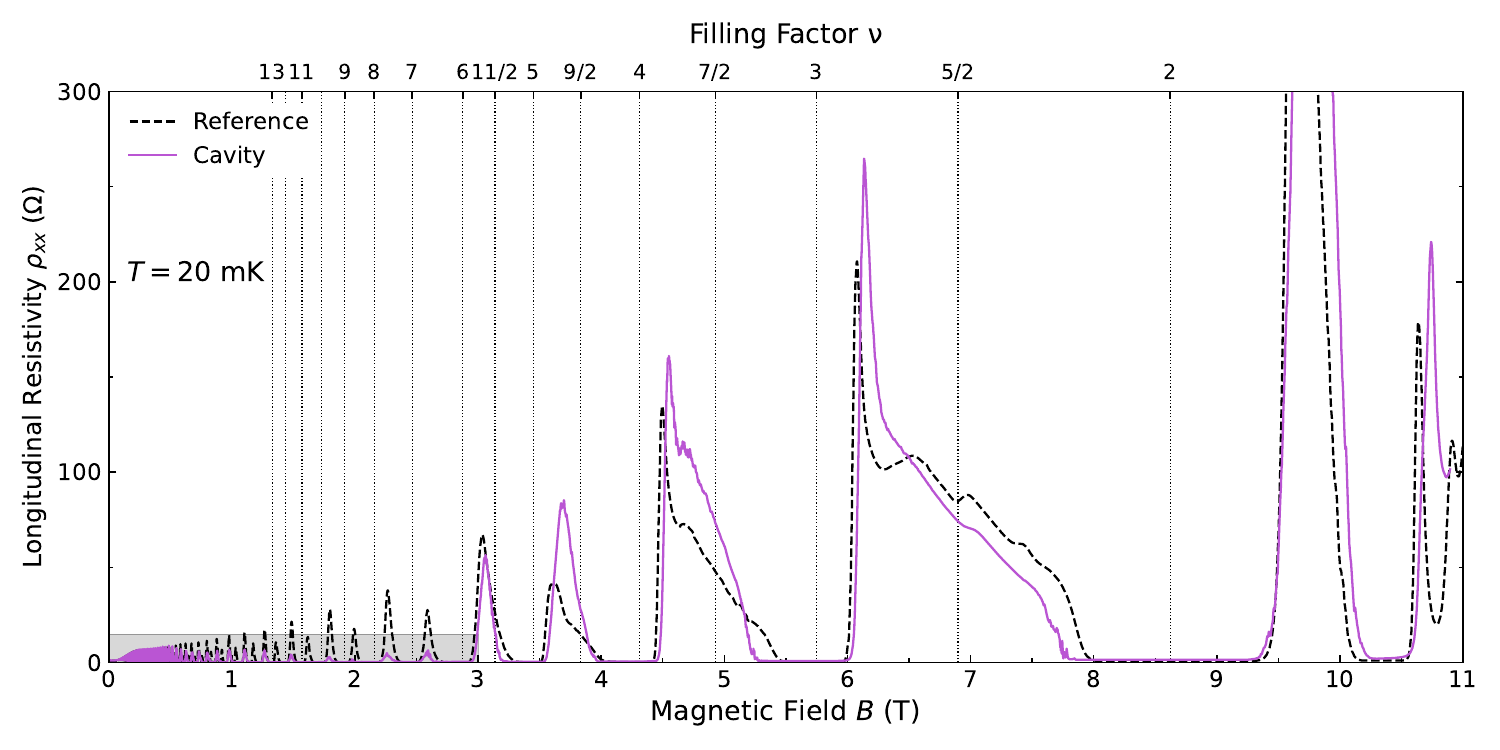}
    \caption{Sample \textbf{D170202B-2}---Longitudinal resistivity as a function of magnetic field, up to \SI{11}{T}, for the reference (black dashed line) and cavity-embedded (purple solid line) HBs. The grey rectangle indicates the magnetic field region up to \SI{3}{T} shown in Fig.~1A in the main text. We notice in particular the absence of anisotropy at filling factor $\nu = 9/2$ and $11/2$.}
    \label{fig:Bfield11T}
\end{figure}

\section{Data at higher magnetic fields}\label{sec:Bfield11T}
In Figure~\ref{fig:Bfield11T} we report the longitudinal resistivity measured in the reference and cavity sample for a larger range of magnetic fields with respect to the one discussed in the main text. We observe how above \SI{3}{T} the impact of the cavity presence on the magnetotransport is much less evident, specifically we do not observe an anisotropy at filling factor $\nu = 9/2$ and $11/2$. The first reports of transport anisotropy~\cite{lilly1999evidence, du1999strongly} focused instead specifically on these half-integer values, although therein the mechanism for generating the anisotropy and the consequent collective macroscopic alignment of the quantum Hall stripes was different from the one we employ in the present work. Since the mechanism for aligning the stripes which we propose in the main text does not take into account~$\nu$, we wonder if the lack of anisotropy at this $\nu$ could reflect an absence even of the microscopic emergence of stripes---which still segregate into domains with different orientations---which could be caused by the presence of the cavity. The lack of anisotropy at $\nu = 7/2$ and below is instead consistent with all experiments reported so far in the literature.


\section{Characterizing the intrinsic material anisotropy}\label{sec:squarepatch}

In this section we present and discuss transport measurements performed on a square patch with \SI{1}{mm} side length, which is physically located on the same chip where the reference and the cavity-embedded Hall bars discussed in the main text are. In this way, the square patch is fabricated concurrently with the Hall bar samples, it comprises a 2DES hosted in exactly the same heterostructure, and it is measured in the same cool-down run with the same measurement scheme. In the following, we show how the intrinsic material anisotropy can account for at most a factor of 1.18 difference in the longitudinal conductivities $\rho_{xx}$ and $\rho_{yy}$ along the two crystallographic directions, two to three orders of magnitude less than the cavity-induced anisotropy we report in the main text, and with the opposite sign.

\begin{figure}
    \centering
    \includegraphics[width=0.5\linewidth]{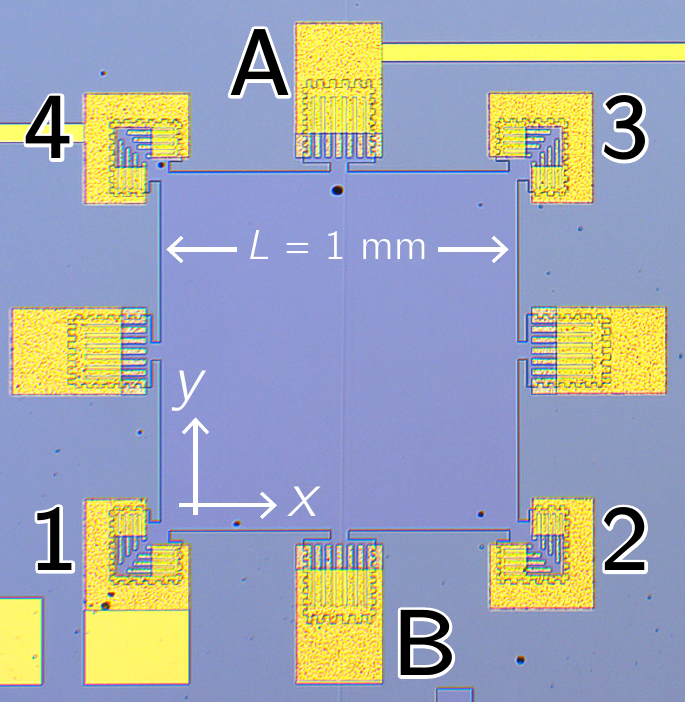}
    \caption{Optical microscope picture of the \SI{1}{mm}-side square patch located on the same chip as the Hall bar samples discussed in the main text. The numbers and letters label the Ohmic contacts nearby (the grainy yellow color comes from the metal annealing process), and we will refer to them in the following discussion. Notice the orientation of the Cartesian axes, which is the same one employed for the reference and cavity-embedded Hall bars.}
    \label{fig:squarepatch_pic}
    \end{figure}

In Figure~\ref{fig:squarepatch_pic} we display an optical microscope picture of the square patch, along with the labeling of the contacts which we refer to in the following measurement plots. Transport is investigated via four-point resistance measurements: a current $I = \SI{10}{nA}$ root-mean-square (rms) is symmetrically injected at contacts $i,j$ (by applying \SI{0.2}{V} rms on two \SI{10}{M \ohm} resistors in front of the contacts, sinusoidally modulated at the lock-in \SI{13.333}{Hz} oscillator frequency), and the voltage difference is measured between contacts $k,l$. Dividing the latter by $I$, we obtain the resistance $R_{ij,kl}$.

\begin{figure}
    \centering
    \includegraphics[width=\linewidth]{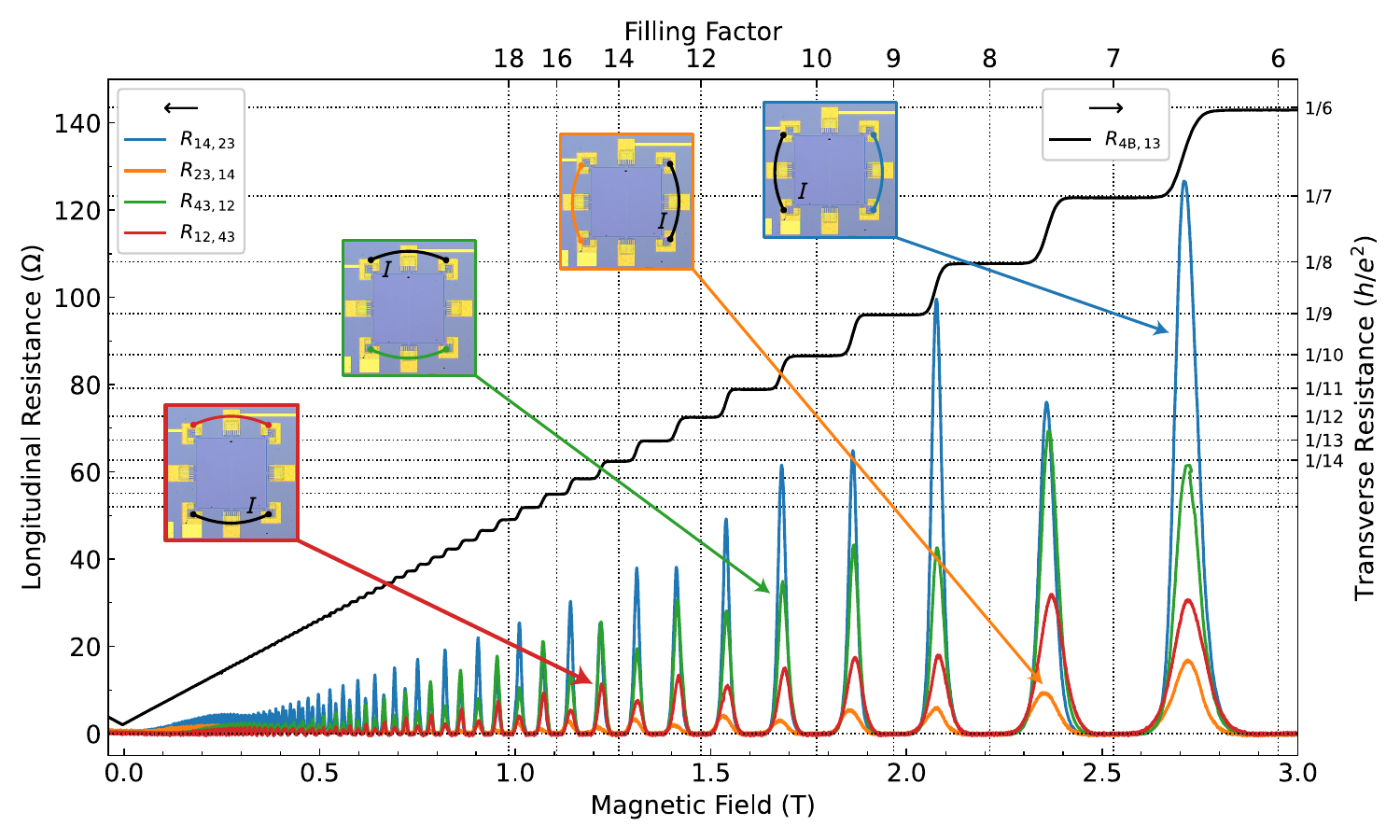}
    \caption{Longitudinal (left axis) and transverse (right axis) resistance of the square patch, as a function of magnetic field, for different current injection and measurement contact pairs, as depicted in the insets. The labeling $R_{ij,kl}$ in the legends employs the same notation introduced in Fig.~\ref{fig:squarepatch_pic}. The top axis shows the filling factor corresponding to the magnetic field shown on the bottom axis.}
    \label{fig:squarepatch_resistances}
\end{figure}

In Figure~\ref{fig:squarepatch_resistances} we report the longitudinal and transverse (Hall) resistance as a function of magnetic flux density (magnetic field) $B$, measured with different current injection and measurement contact pairs, labeled with the notation introduced above and marked for clarity in the insets. Although at some magnetic field values the longitudinal resistance traces differ almost by an order of magnitude, the difference is entirely explained by the presence of a density gradient through the sample, and leads to an almost entirely isotropic resistivity tensor.

To understand the origin of such a difference we elaborate on the model introduced in Ref.~\cite{stern2006densitygradient} to describe current flow in a 2DES with a gradient in the electronic density, assuming, at variance with the treatment therein, different longitudinal resistivities $\rho_{xx}$, $\rho_{yy}$ along the two Cartesian axes. Our model allows us to extract both resistivities employing only the four longitudinal resistance traces shown in Fig.~\ref{fig:squarepatch_resistances}, without any additional adjustable parameter. 

The local electric field $\vb{E}(\vb{r})$ in the two-dimensional sample is given by
\begin{equation}\label{eq:Eresj}
    \vb{E}(\vb{r}) = \begin{pmatrix}
            \rho_{xx} & \rho_{xy} \\
            -\rho_{xy} & \rho_{yy}
            \end{pmatrix} \vb{j}(\vb{r}),
\end{equation}
where $\vb{j}(\vb{r})$ is the local current density, and the components of the resistivity tensor $\rho_{ij}$ are also functions of position, due to inhomogeneities of the electron density in the sample. In the steady-state the following equations must hold
\begin{equation}\label{eq:divj}
    \div{\vb{j}} = 0
\end{equation}
\begin{equation}\label{eq:curlE}
    \curl{\vb{E}} = 0,
\end{equation}
and we can ensure Eq.~\ref{eq:divj} by defining a stream function $\psi$ via 
\begin{equation}
    \vb{j} = \begin{pmatrix}
            -\partial_y \psi \\
            \partial_x \psi
            \end{pmatrix}.
\end{equation}
Substituting Eq.~\ref{eq:Eresj} into Eq.~\ref{eq:curlE}, the equation for $\psi$ is
\begin{equation}\label{eq:psi_full}
    \left(\rho_{yy}\partial_{xx}^2 \psi + \rho_{xx}\partial_{yy}^2 \psi\right) + \left(\partial_x \rho_{yy} \partial_x \psi + \partial_y \rho_{xx} \partial_y \psi \right) + \left( \partial_x \rho_{xy} \partial_y \psi - \partial_y \rho_{xy} \partial_x \psi\right) = 0,
\end{equation}
which, together with the appropriate boundary conditions related to the current injection contacts, can be solved to yield the current distribution, and by integration the voltage drop between any pair of measurement contacts.

\begin{figure}
    \centering
    \includegraphics[width=0.8\linewidth]{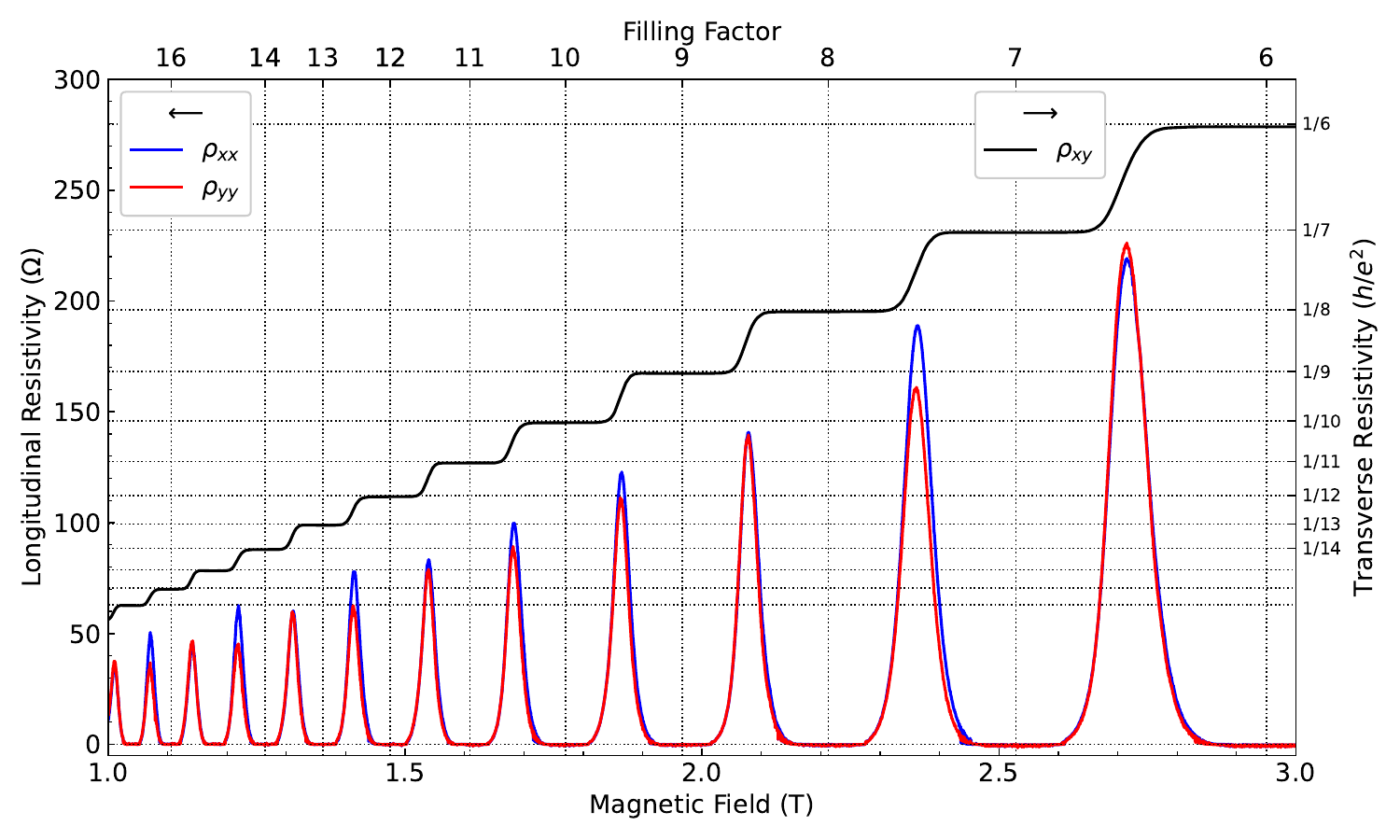}
    \caption{Longitudinal (left axis) and transverse (right axis) resistivity of the square patch, as a function of magnetic field, between \SI{1}{} and \SI{3}{T}. The longitudinal resistivities have been obtained from the longitudinal resistance traces shown in Fig.~\ref{fig:squarepatch_resistances} via the analytical procedure described in the text, while the transverse resistivity coincides with the transverse resistance. The blue and red curves refer to $\rho_{xx}$ and $\rho_{yy}$, respectively.}
    \label{fig:squarepatch_resistivities}
\end{figure}

For magnetic field values in which the 2DES has a density close to a compressible state, and such that $\abs{\rho_{xy}} \gg \rho_{xx}, \rho_{yy}$ (as it is the case for $B > \SI{1}{T}$), we can assume that $\rho_{xx}, \rho_{yy}$ are independent of position, while $\rho_{xy}$ has a constant gradient $\grad{\rho_{xy}}$ in an arbitrary direction. This gradient reflects the density inhomogeneities present in the sample. Thus, Eq.~\ref{eq:psi_full} simplifies to
\begin{equation}\label{eq:psi_simpl}
    \left(\rho_{yy}\partial_{xx}^2 \psi + \rho_{xx}\partial_{yy}^2 \psi\right) + \left( \partial_x \rho_{xy} \partial_y \psi - \partial_y \rho_{xy} \partial_x \psi\right) = 0.
\end{equation}
Without loss of generality, we can assume that current is injected and extracted from contacts $4$ and $3$, and that the voltage difference is measured between contacts $1$ and $2$, following the labeling of Fig.~\ref{fig:squarepatch_pic}. The four configurations employed to measure the longitudinal resistances in Fig.~\ref{fig:squarepatch_resistances} can be retrieved via a suitable rotation of Cartesian axes. According to the position of the injection contacts, the boundary conditions are 
\begin{gather}
        \psi(x,y=0) = \psi(x=0,y) = \psi(x=L,y) = 0 \\
    \psi(x,y=L) = -I,
\end{gather}
and we introduce the following length scales to quantify the carrier density variation across the sample
\begin{equation}
    \ell_x \equiv \rho_{xx}\left(\partial_x \rho_{xy}\right)^{-1},
\end{equation}
\begin{equation}
    \ell_y \equiv \rho_{yy}\left(\partial_y \rho_{xy}\right)^{-1}.
\end{equation}
Using the separation of variables method, the solution to Eq.~\ref{eq:psi_simpl} is
\begin{equation}
    \begin{split}
        \psi = -I\sum_{m=0}^{\infty}\frac{2m\pi\left(1-(-1)^m e^{-L/2\ell_y}\right)}{(L/2\ell_y)^2 + m^2\pi^2} 
        \times \exp\left(\frac{x}{2\ell_y} - \frac{(y-w)}{2\ell_x}\right) 
        \times \frac{\sinh\left(\frac{y}{2}\sqrt{(1/\ell_x)^2+4\lambda_m}\right)}{\sinh\left(\frac{L}{2}\sqrt{(1/\ell_x)^2+4\lambda_m}\right)} \sin\left(\frac{m\pi x}{L}\right),
    \end{split}
\end{equation}
where 
\begin{equation}
    \lambda_m = \frac{\rho_{yy}}{\rho_{xx}}\left(\frac{m^2 \pi^2}{L^2} + \left(\frac{1}{2\ell_y}\right)^2\right).
\end{equation}
We have numerically checked that for the length scales relevant in the experiment ($L = \SI{1000}{\micro\metre}$, $\ell_x, \ell_y \gtrsim \SI{100}{\micro\metre}$) we can retain only the term $m=1$ in the sum without introducing a significant error. The voltage difference between contacts $1$ and $2$ (that is the voltage drop on the opposite side of the one where the current is injected) is given by
\begin{equation}
    V_{12} = \int_0^L dx\,E(x,y=0) = \rho_{xx} \int_0^L dx\,j_x(y=0) = -\rho_{xx} \int_0^L dx\, \partial_y \psi\vert_{y=0},
\end{equation}
and the resistance is $R_{12} = V_{12} / I$. We can now adapt the solution to the different injection and measurement contact pairs, obtaining e.g.\
\begin{equation}\label{eq:R1423}
    R_{14,23} = \rho_{yy} \frac{\frac{1}{2}\sqrt{1/\ell_y^2 + \left(1/\ell_x^2 + 4\pi^2/L^2\right)\frac{\rho_{xx}}{\rho_{yy}}}}{\sinh\left(\frac{L}{2}\sqrt{1/\ell_y^2 + \left(1/\ell_x^2 + 4\pi^2/L^2\right)\frac{\rho_{xx}}{\rho_{yy}}}\right)} \frac{2\pi^2 L}{\left((L/2\ell_x)^2 + \pi^2 \right)^2} \left(1+e^{L/2\ell_x}\right)\left(1+e^{-L/2\ell_x}\right) e^{L/2\ell_y},
\end{equation}
where $\ell_x, \ell_y, \rho_{xx}, \rho_{yy}$ refer to the Cartesian system defined in Fig.~\ref{fig:squarepatch_pic}. All other resistances can be obtained by either changing $\ell_x$ to $\ell_y$, or $\rho_{xx}$ to $\rho_{yy}$, and vice-versa.
From the ratio between the resistances measured on two opposite sides (interchanging injection and measurement contact pairs) we immediately obtain the density inhomogeneity length scales
\begin{align}
        \ell_x &= \frac{L}{\log(R_{43,12}/R_{12,43})}, \\
        \ell_y &= \frac{L}{\log(R_{14,23}/R_{23,14})},
\end{align}
which can then be substituted back in Eq.~\ref{eq:R1423} and the analogous ones, to obtain $\rho_{xx}$ and $\rho_{yy}$. Notice that the four equations for the four longitudinal resistance traces contain only the four unknowns $\ell_x, \ell_y, \rho_{xx}, \rho_{yy}$, so we do not need to know beforehand the density inhomogeneity magnitude, apart from assuming that its gradient is constant. We remind that the transverse resistivity $\rho_{xy}$ is instead simply equal to the transverse resistance $R_{4\mathrm{B},13}$. In Fig.~\ref{fig:squarepatch_resistivities} we report both the longitudinal and transverse resistivities of the square patch, as obtained with the procedure described above. Notice that the 2DES displays at most an 18\% anisotropy: for filling factor $7+1/2$, $\rho_{xx} \simeq 1.18\rho_{yy}$, which is even in the opposite direction with respect to the anisotropy observed in the cavity-embedded Hall bar and reported in the main text, where $\rho_{yy} \gg \rho_{xx}$. The values of $\ell_x, \ell_y$ obtained at the half-integer filling factors range between \SIrange{1}{1.5}{mm} and \SIrange{0.3}{0.5}{mm}, respectively---that is, the density gradient is stronger in the $y$ direction, and the density decreases along the $y$ direction since $\ell_y > 0$---and they are consistent with the density gradient estimated by measuring the Hall resistance on different Hall bars located at different points on the chip. We remark once more that the model we have employed here completely disregards nonlocal effects due to the presence of edge states, and disequilibrium between edge and bulk transport channels.

\begin{figure}
    \centering
    \includegraphics[width=\linewidth]{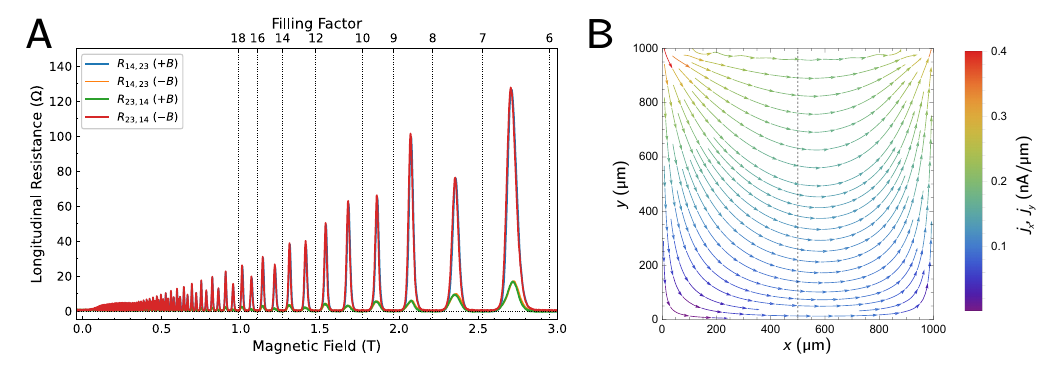}
    \caption{(\textbf{A}) Longitudinal resistance for two pairs of current and voltage probes (following the labelling of Fig.~\ref{fig:squarepatch_pic}) for positive and negative magnetic field direction, as indicated in the legend. Notice how Onsager's relations hold. (\textbf{B}) Current density distribution inside the square sample (when current is injected through contacts 4, 3 as labelled in Fig.~\ref{fig:squarepatch_pic}), as calculated with the model described in the main text, and employing the values of $\ell_x \approx \SI{1.4}{mm}, \ell_y \approx \SI{0.5}{mm}, \rho_{xx} = \rho_{yy} \approx \SI{220}{\ohm}$ obtained experimentally at filling factor $6+1/2$, where the sample is entirely isotropic. Notice the strong decay of the current density magnitude towards the lower end of the sample. The asymmetric profile noticeable when looking along $x=\SI{500}{\micro m}$ (dashed line) is due to the carrier density gradient.}
    \label{fig:panel_supp_ons_curr}
\end{figure}

We have also experimentally verified that Onsager's relations~\cite{buttiker1988symmetry} hold upon the simultaneous interchange of current and measurement contacts and magnetic field reversal, i.e.\
\begin{equation}
    R_{ij,kl}(+B) = R_{kl,ij}(-B),
\end{equation}
such that measurements at negative magnetic field do not add additional information (see Fig.~\ref{fig:panel_supp_ons_curr}A).

We can now comment on the almost entirely isotropic resistivity of our material at variance with what was observed in Ref.~\cite{lilly1999evidence, du1999strongly}. First of all, due to simple geometrical reasons related to the current distribution inside the sample (see Fig.~\ref{fig:panel_supp_ons_curr}B), even a small anisotropy ratio is strongly enhanced by employing a square patch~\cite{simon1999comment}, and special care should be applied when extracting the longitudinal resistivities from the resistance measurements, following the procedure we explained above. According to Ref.~\cite{simon1999comment}, indeed, an anisotropy ratio of $\rho_{yy}/\rho_{xx} \approx 7$ is sufficient to produce the ratio $R_{yy}/R_{xx} \approx 60$ observed by Lilly \emph{et al.}~\cite{lilly1999evidence} in a square sample. Secondly, it has been shown that the stripe orientation can be influenced by strain~\cite{koduvayur2011effect} or by patterning an artificial one-dimensional charge modulation on the surface of the 2DEG~\cite{willett2001current}. We are therefore led to attribute the lack of anisotropy in our material to an especially smooth sample surface and to the high quality of the epitaxial growth.

\begin{figure}
    \centering
    \includegraphics[width=\linewidth]{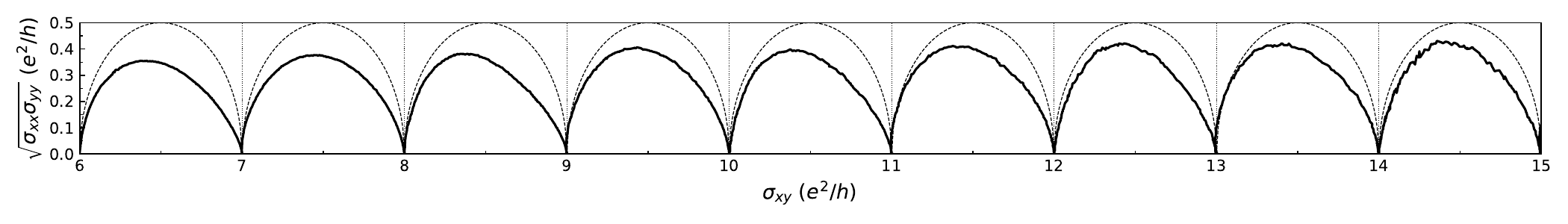}
    \caption{Square root of the product of the longitudinal conductivities, as a function of the transverse conductivity (solid line), as obtained from inverting the resistivity tensor. Notice how the data closely follow the semicircle relation (dashed line): lack of perfect agreement can be attributed to finite-size effects.}
    \label{fig:mceuen_square}
\end{figure}

\begin{figure}
    \centering
    \includegraphics[width=0.4\linewidth]{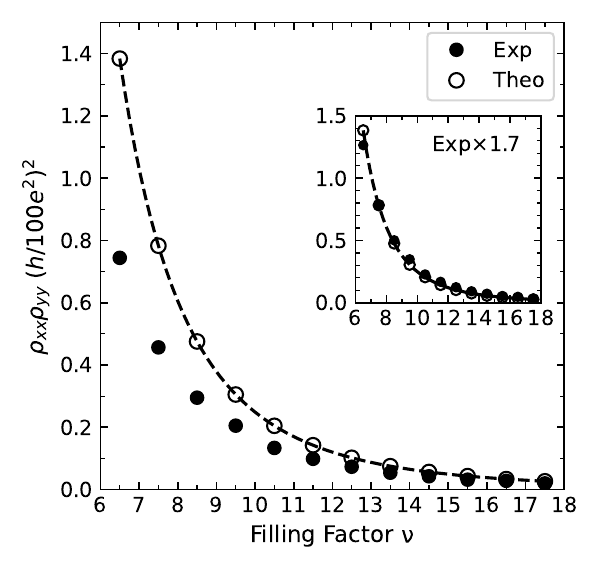}
    \caption{Product of longitudinal resistivities at half-integer filling factors, as obtained from the experimental data of Fig.~\ref{fig:squarepatch_resistivities} (solid circles) or from the universal relationship of Eq.~\ref{eq:product_resistivities}. Lack of agreement comes from finite-size effects: in the inset it is shown that simply multiplying the experimental data by 1.7 recovers almost exactly the theoretical prediction.}
    \label{fig:product_res_sq}
\end{figure}

Finally, inverting the resistivity tensor one obtains the conductivity tensor
\begin{equation}
    \begin{pmatrix}
        \sigma_{xx} & -\sigma_{xy} \\
        \sigma_{xy} & \sigma_{yy}
    \end{pmatrix} = \begin{pmatrix}
        \rho_{xx} & \rho_{xy} \\
        -\rho_{xy} & \rho_{yy}
    \end{pmatrix}^{-1} = \frac{1}{\rho_{xx}\rho_{yy}+\rho_{xy}^2}\begin{pmatrix}
        \rho_{yy} & -\rho_{xy} \\
        \rho_{xy} & \rho_{xx}
    \end{pmatrix},
\end{equation}
and we can thus verify that the sample fulfills the semicircle relation~\cite{dykhne1994theory} adapted to the case of a generally anisotropic tensor~\cite{von2000conductivity}:
\begin{equation}
    \sigma_{xx}\sigma_{yy} + \left(\sigma_{xy} - (N + 1/2)e^2/h \right)^2 = (e^2/2h)^2,
\end{equation}
where $N$ is a non-negative integer. In Fig.~\ref{fig:mceuen_square} we show $\sqrt{\sigma_{xx}\sigma_{yy}}$ as a function of $\sigma_{xy}$, which closely follows the semicircle relation (shown with the dashed line). Lack of perfect agreement comes from the finite size of the $\SI{1}{mm}\times\SI{1}{mm}$ square patch. Size effects are also manifest when comparing (Fig.~\ref{fig:product_res_sq}) the product of the longitudinal resistivities, which according to Ref.~\cite{sammon2019resistivity} should be given, at half-integer filling factors $\nu$ and even in the presence of anisotropy, by
\begin{equation}\label{eq:product_resistivities}
    \rho_{xx}\rho_{yy} = \left(\frac{h}{e^2}\right)^2 \frac{1}{\left(2\nu^2 + 1/2\right)^2}.
\end{equation}
The experimental data superimpose to the universal behaviour when they are multiplied by a factor of $1.7$.

\section{Longitudinal resistivity in the presence of a density gradient}\label{sec:density_inhom}

\begin{figure}
    \centering
    \includegraphics[width=\linewidth]{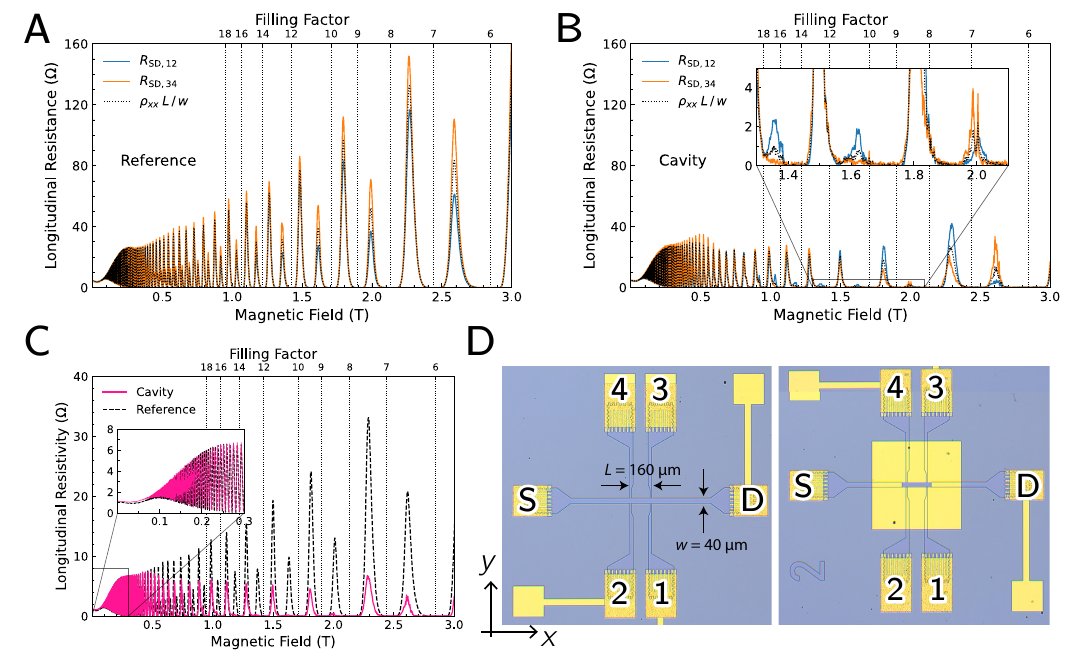}
    \caption{Sample \textbf{D170202B-2}---(\textbf{A}) Longitudinal resistance measured on the two sides of the reference Hall bar, by passing currrent through the source-drain contacts (S-D in (D)) and employing voltage probes 1-2 or 3-4 (as labelled in (D)). The dotted line is the longitudinal resistivity (still multiplied by the geometric factor $L/w$) obtained by combining the measurements on the two sides with the procedure explained in the text. Notice that at all magnetic fields it lies in between them. (\textbf{B}) Longitudinal resistance measured on the two sides of the cavity-embedded Hall bar, with the same measurement scheme of (A). In the inset an enlargement of the region where the cavity suppresses the resistivity is displayed. (\textbf{C}) Longitudinal resistivity $\rho_{xx}$ compared between cavity-embedded and reference sample. The curves are the same dotted ones of (A, B), divided by the $L/w = 4$. (\textbf{D}) Optical microscope pictures of the reference and cavity-embedded Hall bars, with the labelling of the contacts.}
    \label{fig:longitudinal_density}
\end{figure}

In Section~\ref{sec:squarepatch} we have quantified the inhomogeneous density $n_s$ of the 2DES, and shown that its gradient is related to the length scale of the variation of the Hall resistance, which amounts to about \SI{1.2}{mm} along the $x$ direction and \SI{0.4}{mm} along the $y$ direction. Although the Hall bars whose measurements we report in the main text have $w=\SI{40}{\micro m}$ width and $L=\SI{160}{\micro m}$ length, one order of magnitude smaller than the scale density modulation, which is then usually overlooked in usual quantum Hall measurements, we have applied great care in obtaining the longitudinal resistivity since, being the claimed stripe-ordered phase a charge density modulation, we want to avoid being misled by spurious effects due to other density inhomogeneities. In Fig.~\ref{fig:longitudinal_density}A,B we report the longitudinal resistance measured at positive magnetic fields on the two sides of both the reference and cavity-embedded Hall bar, $R_{\mathrm{SD},12}, R_{\mathrm{SD},34}$ (where the first two subscripts refer to the injection contacts, and the second two to the voltage probes, according to the labelling of Fig.~\ref{fig:longitudinal_density}D). It is clear that the two sides do not superimpose, the reason being due to density inhomogeneities, which at magnetic fields above \SI{1}{T} can impact the current distribution across the sample. As noted in Ref.~\cite{stern2006densitygradient}, even a slight density gradient along the Hall bar leads to differences in Hall resistances $R_{\mathrm{SD},24}, R_{\mathrm{SD},13}$, which reflect in differences of $R_{\mathrm{SD},12}, R_{\mathrm{SD},34}$. Indeed, if $V_2 - V_4 > V_1 - V_3$ (a consequence of having a larger $n_s$ in the region between probes 1-3 with respect to 2-4, since the Hall resistance is inversely proportional to $n_s$, away from plateaux), it immediately follows $V_1 - V_2 < V_3 - V_4$, and also $\Delta R_\mathrm{long} = R_{\mathrm{SD},34} - R_{\mathrm{SD},12} = R_{\mathrm{SD},24} - R_{\mathrm{SD},13}$. For example, at \SI{1}{T} and with $n_s = \SI{4e11}{cm^{-2}}$ a difference of 1\% in density reflects in $\Delta R_\mathrm{long} \approx \SI{15}{\ohm}$. We have checked that the previous equality holds for both cavity-embedded and reference samples, which also amounts to a sanity check on our measurement setup. 

Following Ref.~\cite{stern2006densitygradient}, it is straightforward to extract the longitudinal resistivity $\rho_{xx}$, even when the density gradient is unknown, via the following equation
\begin{equation}\label{eq:rho_extract}
    \rho_{xx} = \frac{w}{L}\frac{R_{\mathrm{SD},34} - R_{\mathrm{SD},12}}{\log\left(R_{\mathrm{SD},34}/R_{\mathrm{SD},12}\right)},
\end{equation}
a result which follows from the same electrostatic model discussed in Sec.~\ref{sec:squarepatch}, and which stays true also when $\rho_{xx} \neq \rho_{yy}$ (the anisotropic case which we are most interested to). We report the extracted resistivity in Fig.~\ref{fig:longitudinal_density}A-C (in A,B it is still multiplied by the geometric factor $L/w = 4$). Finally, reversing the direction of the magnetic field simply changes the sign of the resistance difference, and it does not affect the value given by Eq.~\ref{eq:rho_extract}.

In complete analogy, we have combined the two measured nonlocal resistances $R_{24,13}, R_{13,24}$ to obtain the nonlocal resistance we report in the main text
\begin{equation}\label{eq:RNL_extract}
    R_{yy} = \frac{R_{24,13} - R_{13,24}}{\log\left(R_{24,13} / R_{13,24}\right)}.
\end{equation}

Finally, the transverse resistances reported in the main text are just the average of $R_{\mathrm{SD},24}, R_{\mathrm{SD},13}$ or of $R_{24,\mathrm{SD}},R_{13,\mathrm{SD}}$, for $\rho_{xy}, \rho_{yx}$, respectively.


\section{Nonlocal resistance and edge states transport model}
The longitudinal \textit{resistivity} along the $\hat{\vb{x}}$ direction, as presented in Fig~1A in the main text, is simply obtained via $\rho_{xx} = R_{xx} / 4$---given by the ratio between the width $w$ and the length $L$ (i.e.\ the distance between voltage probes) of the HB (see Fig.~1C in the main text). In contrast, however, the resistance $R_{yy}$, especially in small devices like the HB that we use here, cannot be interpreted simply as a \textit{bulk} $\rho_{yy}$. In particular $R_{yy}$ not only includes contributions from the bulk resistivity in the $\hat{\vb{y}}$ direction (i.e.\ across the stripes in the cavity sample) but also nonlocal contributions from edge states which, in fact, dominate the bulk contributions.  Indeed, if one were to simply apply Ohm's law to describe the classical current flow in the HB, one would obtain that the value of $R_{yy}$ would be $\sim 4/\pi \, \rho_{xx} \exp(-\pi L / w) \sim 4\times10^{-6}\rho_{xx}$~\cite{abanin2011giant}. The fact that instead we do observe a much larger value of $R_{yy}$ stems from the presence of edge states---a characteristic of the quantum Hall regime---and it is thus termed \emph{nonlocal} resistance~\cite{wang1992measurements}. We emphasize that the attribution of $R_{yy}$ to a measure of nonlocal resistance follows directly from the geometry of our HB and is, crucially, independent of the cavity.  


In this section we discuss the model to characterize transport at half-integer filling factors, which allows us to obtain from the longitudinal and nonlocal resistances an estimate of the scattering amplitudes into different transport quantum channels. The model is proposed in Ref.~\cite{mceuen1990new}, and assumes that for filling factors $\nu \in [N-1,N)$, with $N$ a positive integer, transport occurs via $N-1$ dissipationless edge states and a single dissipative bulk channel, decoupled from the other channels, and corresponding to the partially backscattered innermost edge state in the $N$-th spin-resolved Landau level. We employed the same model in Ref.~\cite{appugliese2022breakdown}, albeit therein to study the influence of vacuum fields on transport at integer filling factors, in particular by positing that vacuum field-induced backscattering destroys the quantization and gives a finite resistance to the otherwise dissipationless quantum Hall states. Moreover, due to the usage of a more compact Hall bar geometry in the present work, the model can be easily employed to compare the estimated nonlocal resistance with the experimental measurements.

\begin{figure}
    \centering
    \includegraphics[width=0.9\linewidth]{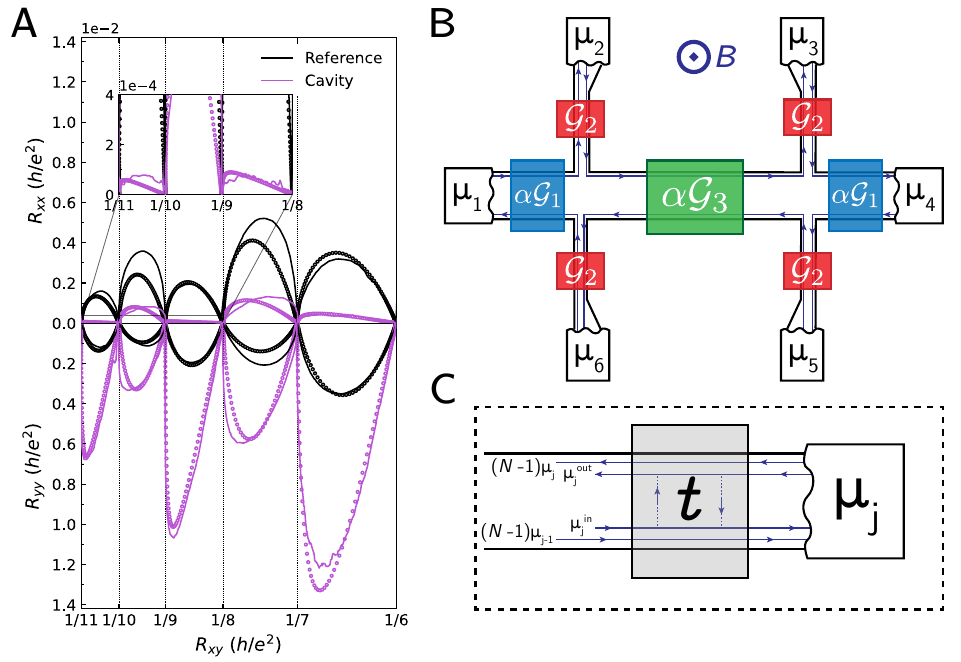}
    \caption{(\textbf{A}) Longitudinal (top panel) and nonlocal resistance (bottom panel, with inverted axis)  as a function of transverse resistance, for the reference (black solid curve) and cavity-embedded Hall bar (purple solid curve). The black and purple empty circles show the result obtained from the edge state transport model, for the reference and cavity sample, respectively. The inset shows an enlargement between filling factors 8 and 11, where we observe the cavity-induced strong suppression of longitudinal resistance. (\textbf{B}) Edge state transport model, showing the geometrical factors $\mathcal{G}_i$ related to the different sections of the Hall bar. We multiply the geometrical factors $\mathcal{G}_1$ and $\mathcal{G}_3$ by a factor $\alpha \ll 1$ to take into account the presence of the stripe-ordered phase, which reduces the probability of scattering from one side to the other. For the isotropic liquid phase $\alpha=1$. (\textbf{C}) Sketch of the scattering process which connects edge states going in opposite directions on the two sides of a section of the Hall bar. While $N-1$ edge states are completely transmitted, the innermost edge state corresponding to the topmost $N$-th Landau level is partially transmitted, with transmission amplitude $t$, and partially scattered to the other side, with amplitude $1-t$. The chemical potentials $\mu_j, \mu_{j-1}, \mu_j^{\mathrm{in}}, \mu_j^{\mathrm(out)}$ which we refer to in the text are indicated.}
    \label{fig:McEuen_HB_stripes}
\end{figure}

In Fig.~\ref{fig:McEuen_HB_stripes}B we show the schematic employed to define the model: the Hall bar is divided into segments which have geometrical factors $\mathcal{G}$ given by the ratio of their length and width ($\mathcal{G}_1 = 11.5$, $\mathcal{G}_3 = 4$, and $\mathcal{G}_2$ is adjusted to fit the reference data, as detailed below), and which transmit the $N$-th edge state with probability amplitude $t_i$, related to the geometrical factor by
\begin{equation}\label{eq:trasm_ampl}
    t_i = \frac{1}{1+\rho \mathcal{G}_i},
\end{equation}
where $\rho$ is the \emph{resistivity} of the $N$-th channel only, which is made to vary between zero to infinity as a parameter. In particular, when $\rho$ is zero, the $N$-th channel behaves as a dissipationless edge channel, and as it increases to infinity the channel is more and more backscattered---with amplitude $1-t$---while the transverse resistance increases from $(1/N)h/e^2$ to $[1/(N-1)]h/e^2$~\cite{szafer1991network}. 

Applying Büttiker's multiprobe formula~\cite{buttiker1988absence} we get the current injected at each contact $j\in[1,6]$
\begin{equation}
    I_j = \frac{e}{h}\left[(N-1)(\mu_j - \mu_{j-1}) - \mu_j^\mathrm{in} + \mu_j^\mathrm{out}\right],
\end{equation}
with $\mu_j^\mathrm{in}, \mu_j^\mathrm{out}$ being the chemical potentials of the innermost channel only (see Fig.~\ref{fig:McEuen_HB_stripes}C), and $\mu_0 \equiv \mu_6$. From current conservation, we have also
\begin{equation}
    \mu_j^\mathrm{out} = \mu_j t_j + \mu_j^\mathrm{in}(1-t_j),
\end{equation}
as can be checked in Fig.~\ref{fig:McEuen_HB_stripes}C. Finally, $\mu_j^\mathrm{in} = \mu_{j-1}^\mathrm{out}$. These $3\times6=18$ equations in the $18$ variables $\mu_j,\mu_j^{\mathrm{in}},\mu_j^{\mathrm{out}}$ can be solved after providing as boundary conditions the current injected in the leads $I_j$ (for example, to measure the longitudinal resistance we inject current $I$ through contacts 1 and 4, so $I_1 = -I_4 = I$), and we have to fix one value of the chemical potentials as a reference (e.g.\ $\mu_4=0$). After solving the linear system, we can directly obtain the longitudinal, nonlocal, and transverse resistances, and we compare them to the experimental data in Fig.~\ref{fig:McEuen_HB_stripes}A (the experimental data have been combined to remove effects of density inhomogeneities as detailed in Section~\ref{sec:density_inhom}). To model the transport in the stripe-ordered phase, we multiply the geometrical factors $\mathcal{G}_1, \mathcal{G}_3$ by a factor $\alpha < 1$, intended to represent the preferential direction of transport along rather than orthogonal to the stripes ($\alpha=1$ in the isotropic liquid phase of the reference sample). In Table~\ref{tab:McEuen_HB_params} we report the parameters employed to fit the model to the experimental data measured on the cavity sample. The geometrical factor $\mathcal{G}_2$ was fitted first to the experimental data of the reference sample and then not changed in the fitting of the cavity one. In the table we report also the obtained values of the transmission amplitudes, as given by Eq.~\ref{eq:trasm_ampl}, for the reference sample in the isotropic liquid phase ($t_l$ obtained using $\mathcal{G}_3$, with $\alpha=1$), and the cavity sample in the stripe-ordered phase ($t_s$ obtained using $\mathcal{G}_3$, multiplied by $\alpha$ detailed in the table). Considering the backscattering amplitudes, we obtain values of $1-t_s$ one order of magnitude smaller than $1-t_l$ at filling factors $2N+1/2$.

\begin{table}[!hbt]
    \centering
    \scalebox{0.9}{\begin{tabular}{>{\centering\arraybackslash}p{1.5cm}>{\centering\arraybackslash}p{1.5cm}>{\centering\arraybackslash}p{1.5cm}|>{\centering\arraybackslash}p{1.5cm}>{\centering\arraybackslash}p{1.5cm}}
    \toprule
       $\nu$  & $\mathcal{G}_2$ & $\alpha$ & $t_l$ & $t_s$ \\
       \midrule
        6+1/2 & 18 & 0.04 & 0.75 & 0.94\\
        7+1/2 & 6 & 0.07 & 0.75 & 0.88\\
        8+1/2 & 18 & 0.01 & 0.75 & 0.97\\
        9+1/2 & 7 & 0.10 & 0.75 & 0.87\\ 
        10+1/2 & 18 & 0.01 & 0.75 & 0.97\\
        \bottomrule
    \end{tabular}}
    \caption{Parameters employed to fit the model to the experimental data for the cavity sample, and values obtained for the transmission amplitude in the isotropic liquid and stripe-ordered phases ($t_l, t_s$, respectively).}
    \label{tab:McEuen_HB_params}
\end{table}


\section{Experimental reproducibility}

\subsection{Reproducibility across different magnetic field sweeps}
\begin{figure}[!h]
    \centering
    \includegraphics[width=0.8\linewidth]{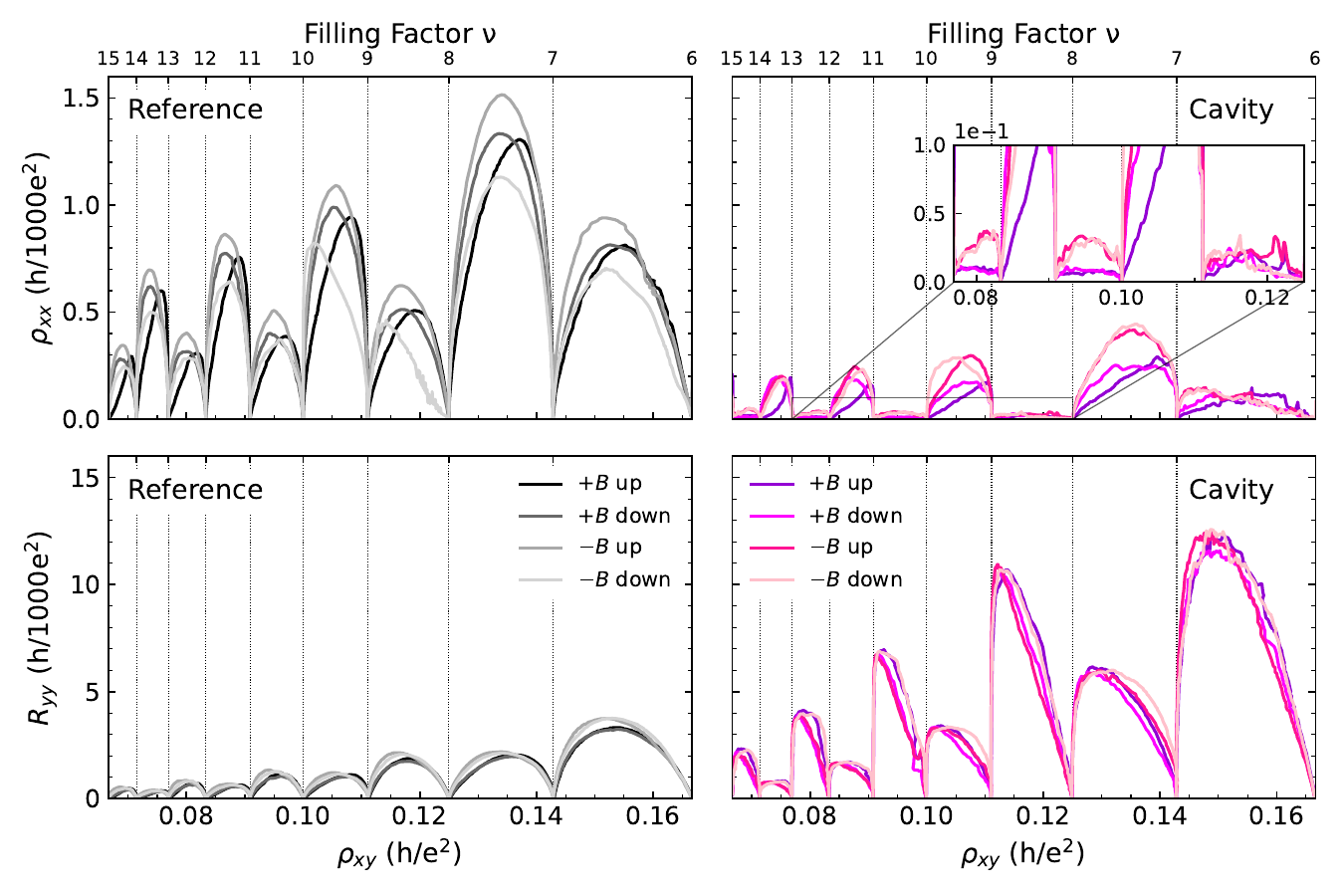}
    \caption{Sample \textbf{D170202B-2}---Longitudinal resistivity $\rho_{xx}$ (top panels) and nonlocal resistance $R_{yy}$ (bottom panels, notice the different $y$-axis scale) as a function of the transverse resistivity $\rho_{xy}$ for the reference (left panels) and cavity-embedded (right panels) Hall bars, for different magnetic field sweeps, that is going from 0 to \SI{3}{T} ($+B$ up), from 3 to \SI{0}{T} ($+B$ down), from 0 to \SI{-3}{T} ($-B$ up), and from $-3$ to \SI{0}{T} ($-B$ down), displayed with colors according to the legends.
    }
    \label{fig:repr_Bsweeps}
\end{figure}

As stated in the main text, we have measured both reference and cavity-embedded Hall bars in the same cool-down, and simultaneously during several magnetic field sweeps, employing multiple synchronized lock-in amplifiers. In particular, we show in Fig.~\ref{fig:repr_Bsweeps} the measurements for 4 successive sweeps at positive and negative magnetic field, and sweeping the magnetic field from zero to its maximum value or vice-versa. We display the results by showing the longitudinal resistivity $\rho_{xx}$ and nonlocal resistance $R_{yy}$ as a function of transverse resistivity $\rho_{xy}$, to better highlight the variations away from the quantized plateaux. We notice that across different sweeps the curves compare fairly well. A plethora of details can still be noticed, such as the slanted shape of the peaks, which seems to partially depend on the sign of the field and direction of the sweep, but this analysis goes far beyond the purpose of the present work.

\subsection{Measurement on a rotated cavity-embedded Hall bar}

\begin{figure}[hbt]
    \centering
    \includegraphics[width=0.9\linewidth]{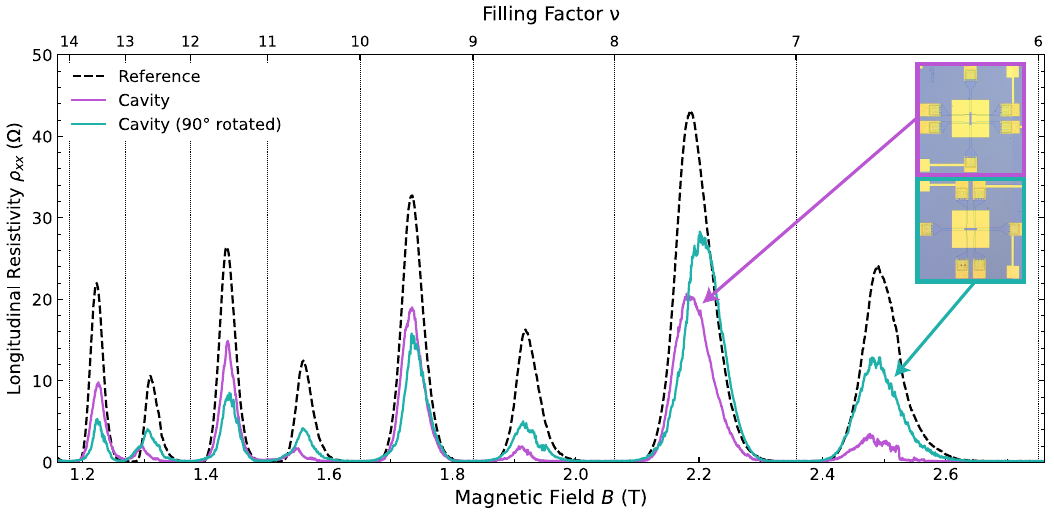}
    \caption{Sample \textbf{D170202B-1}---Longitudinal resistivity as a function of magnetic field for a reference HB (black dashed line), and for two cavity-embedded HBs fabricated on the same chip but rotated by \SI{90}{\degree} with respect to each other (colored solid lines). The inset shows an optical microscope picture of the chip, showing the orientation of the two cavity samples. We observe a similar longitudinal resistivity suppression within the cavity samples---with respect to the reference one---which supports the fact that it is the cavity field orientation to define the collective alignment axis of the stripe order.}
    \label{fig:cav_vert_hor}
\end{figure}

In Figure~\ref{fig:cav_vert_hor} we compare the longitudinal resistivity measured on two cavity-embedded Hall bars fabricated on the same chip (Sample \textbf{D170202B-1}, see Section~\ref{subsubsec:D170202B-1}) but rotated by \SI{90}{\degree} with respect to each other. This is done as a further confirmation that the effect follows the polarization of the cavity vacuum field. Indeed, we observe a similar resistivity suppression in the two cavity samples, as compared to the reference one. The difference in the magnitude of the resistivity suppression comes from the different profile of the cavity edges, as discussed in Section~\ref{sec:cavity_edges}.

\subsection{Role of the cavity edges}\label{sec:cavity_edges}
\begin{figure}[hbt]
    \centering
    \includegraphics[width=\linewidth]{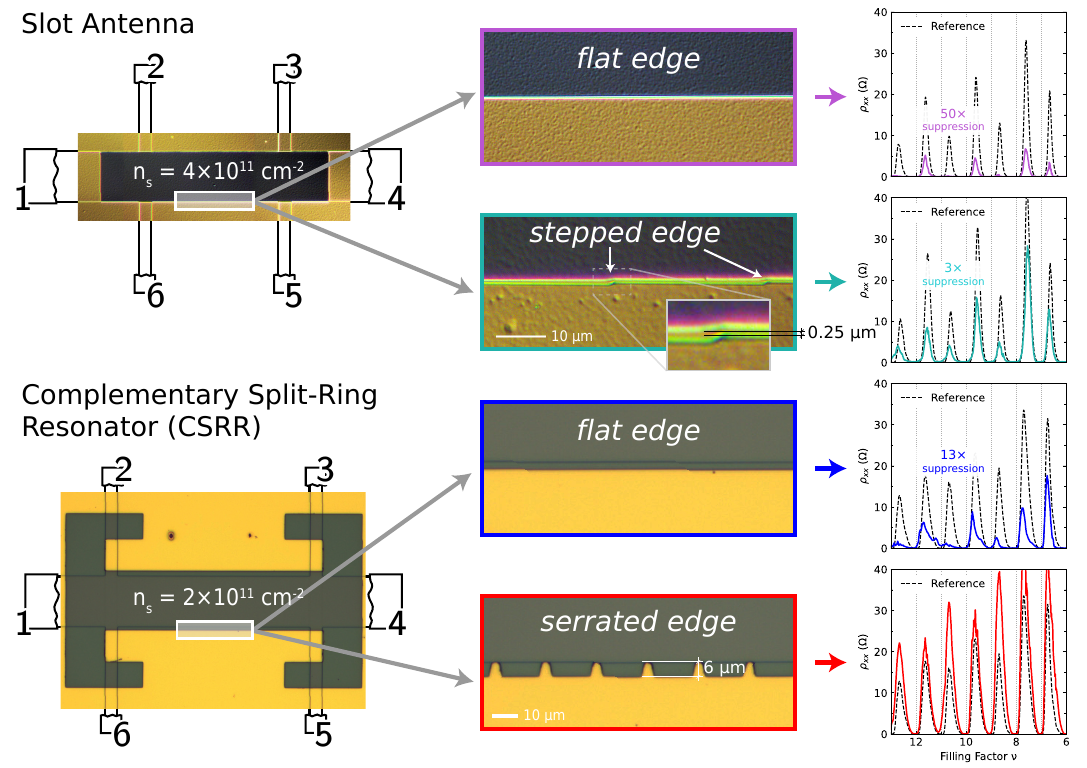}
    \caption{Impact of the cavity edges flatness on the magnitude of the longitudinal resistivity suppression in a cavity-embedded Hall bar. Two different designs of the cavity were employed: the slot antenna which is the object of the present work (top), and the complementary split-ring resonator which was employed in Refs.~\cite{appugliese2022breakdown, enkner2024enhanced} (bottom). For each cavity, two variations have been fabricated, one with flat edges and one with stepped or serrated edges. The four panels on the right report the longitudinal resistivity as a function of filling factor for the respective cavity (colored solid lines) and reference samples (black dashed lines). Notice that the 2DESs have different densities, as indicated on the optical microscope pictures.}
    \label{fig:role_edges}
\end{figure}

In Figure~\ref{fig:role_edges} we explain the effect of the cavity edges on the magnitude of the measured magnetotransport anisotropy. In particular, we considered two cavity designs: the slot antenna already discussed in the main text, and the complementary split-ring resonator (CSRR) which was employed in Refs.~\cite{appugliese2022breakdown, enkner2024enhanced}. The two designs are fabricated on different chips, where the 2DES have densities \SI{4e11}{cm^{-2}}, and \SI{2e11}{cm^{-2}}, respectively, and have been realized in two variations, one with flat cavity edges and one with stepped or serrated edges. On each chip a reference sample is also present. By comparing the different longitudinal resistivities, we observe that the flat edges provide the largest resistivity suppression in the cavity-embedded samples, with a larger suppression in the sample having higher 2DES density. When employing instead a cavity with serrated edges, we clearly see that the resistivity suppression is absent. We did not attempt to quantify the effect further or to relate it to the particular spatial profile of the electromagnetic field mode, which could consitute a topic for further research.

\clearpage

\subsection{Reproducibility across different samples}
In this section we report measurements on other samples processed also on different heterostructures, where we observed a magnetotransport anisotropy comparable to the one reported in the main text. The indication inside the parentheses is the internal labelling of the fabricated samples.

\subsubsection{Second sample processed on D170202B (D170202B-1)}\label{subsubsec:D170202B-1}
\begin{figure}[!h]
    \centering
    \includegraphics[width=\linewidth]{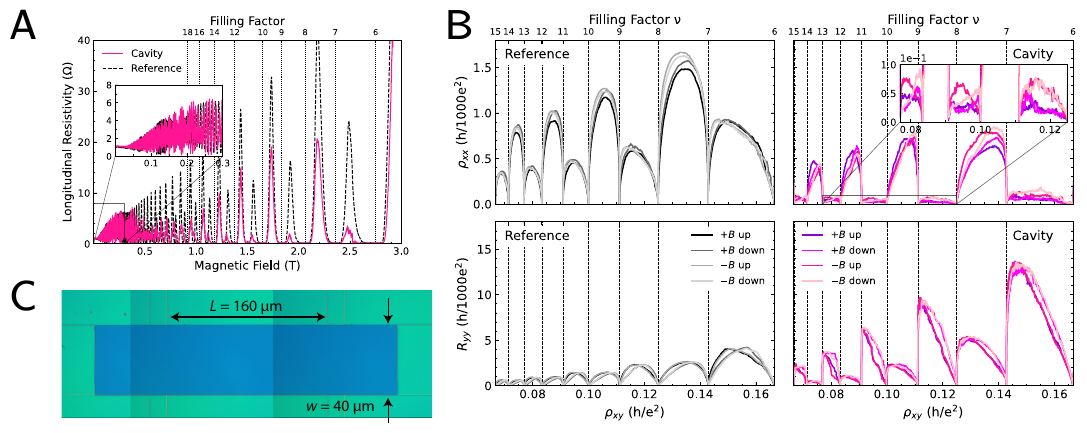}
    \caption{Sample \textbf{D170202B-1}---(\textbf{A}) Longitudinal resistivity calculated for a reference and a cavity-embedded Hall bar, as a function of magnetic field. Around filling factor $8+1/2$ we observe a ratio of about $9.5$. (\textbf{B}) Longitudinal resistivity $\rho_{xx}$ (top panels) and nonlocal resistivity $R_{yy}$ (bottom panels, notice the different y-axis scale) as a function of the transverse resistivity $\rho_{xy}$ for the reference (left panels) and cavity-embedded (right panels) Hall bars, for different magnetic field sweeps, that is going from 0 to \SI{3}{T} ($+B$ up), from 3 to \SI{0}{T} ($+B$ down), from 0 to \SI{-3}{T} ($-B$ up), and from $-3$ to \SI{0}{T} ($-B$ down), displayed with colors according to the legends. Notice how across different sweeps the results are fairly comparable. (\textbf{C}) Optical microscope picture of the slot antenna resonator.}
    \label{fig:D170202B-1-results}
\end{figure}

We fabricated the sample presented in the main text and this one on the same chip, and then cleaved it afterwards. The only difference thus comes from the slightly lower density gradient, due to the different physical position on the heterostructure wafer (the photolitography mask is identical).
In Fig.~\ref{fig:D170202B-1-results} we report the comparison between longitudinal resistivities of the reference and cavity-embedded Hall bars, showing clearly the cavity-suppressed resistivity at half-integer filling factors $2N+1/2$, with $N$ ranging from 3 to 7. We also report the reproducibility study over different magnetic field sweeps (changing sign and ramp direction), assessing again the robustness of the cavity-induced magnetotransport anisotropy.

\subsubsection{Sample processed on D170209B (D170209B-1-40B)}
This sample was one of the very first where the suppression of resistivity at half-integer filling factors in the cavity-embedded Hall bar was observed. It was processed and first measured in 2018, and then remeasured in 2023, with a vastly updated experimental setup: we changed the superconducting Helmholtz coils magnet, capable of reaching \SI{6}{T}, with a solenoidal superconducting magnet, capable of reaching \SI{12}{T}; we remade all electrical wiring and changed chip socket; we changed the capacitors of the \SI{100}{kHz} low-pass filter from \SI{10}{} to \SI{1}{nF}, and increased accordingly the resistors from \SI{1}{k\ohm} to \SI{10}{k\ohm}. Moreover, the 2018 measurement is performed with asymmetrically injected \SI{3}{nA} current, while in 2023 \SI{10}{nA} where injected symmetrically by employing two \SI{100}{M\ohm} resistors before both source and drain. The heterostructure contains a \SI{27}{nm}-wide square quantum well, with a \SI{60}{nm} spacer from the doping layer, and the 2DES has density \SI{4.2e11}{cm^{-2}} and mobility \SI{2.5e7}{cm^2V^{-1}s^{-1}} (measured at \SI{1.3}{K} without illumination). The Hall bar has still a width of \SI{40}{\micro m}, and the voltage probes are \SI{160}{\micro m} apart, however, the overall length is about \SI{1.4}{mm} (since it was probed at different positions), making it harder to model and understand nonlocal measurements. 

\begin{figure}[!h]
    \centering
    \includegraphics[width=\linewidth]{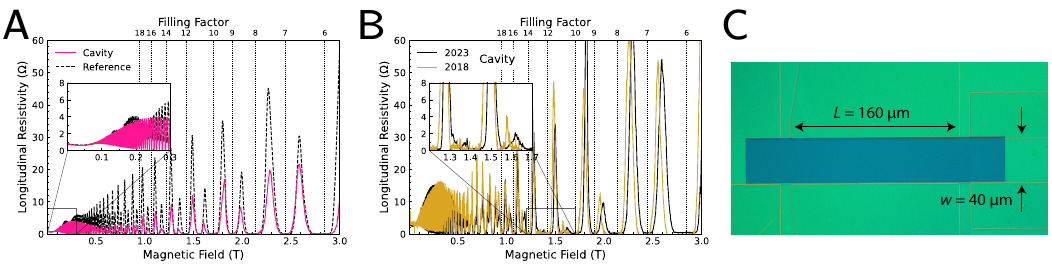}
    \caption{Sample \textbf{D170209B-1-40B}---(\textbf{A}) Longitudinal resistivity calculated for a reference and a cavity-embedded Hall bar, as a function of magnetic field. Around filling factor $12+1/2$ we observe a ratio of about 8. (\textbf{B}) Comparison between the longitudinal resistivity (obtained by taking into account only one side of the Hall bar) of the cavity-embedded sample as measured in 2018 and in 2023, with a vastly updated experimental setup. (\textbf{C}) Optical microscope picture of the slot antenna resonator.}
    \label{fig:D170209B-results}
\end{figure}

In Fig.~\ref{fig:D170209B-results} we report the longitudinal resistivity as a function of magnetic field, where we can clearly distinguish the resistivity suppression in the cavity sample for half-integer filling factors between $10+1/2$ and $14+1/2$. We also display a plot comparing the longitudinal resistivity as measured with the old setup in 2018, and with the new one in 2023, showing perfect agreement.

\subsubsection{Sample processed on D200923 (D200923-Scs\#1)}
\begin{figure}[!h]
    \centering
    \includegraphics[width=\linewidth]{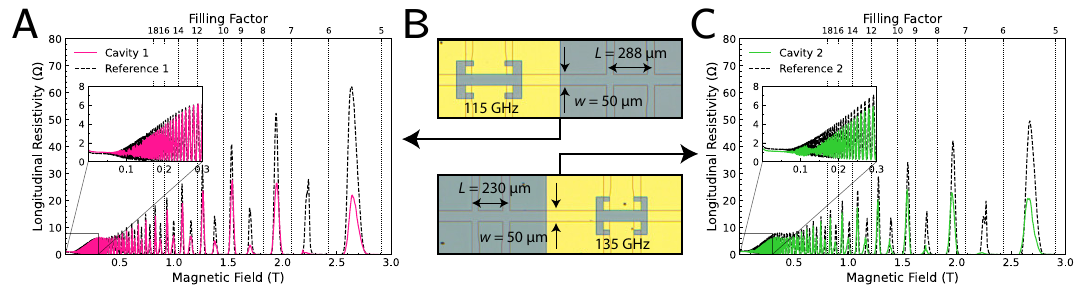}
    \caption{Sample \textbf{D200923-Scs\#1}---(\textbf{A}) Longitudinal resistivity calculated for a reference and a \SI{115}{GHz} cavity-embedded Hall bar, as a function of magnetic field. Around filling factor $6+1/2$ we observe a ratio of about 21. (\textbf{B}) Optical microscope pictures of the two pairs of reference and cavity-embedded resonators. The first pair (top, data are presented in (A)) has distance between voltage probes of \SI{288}{\micro m}, while the second one (bottom, data are presented in (C)) has distance between voltage probes of \SI{230}{\micro m}. Both Hall bar pairs have width \SI{50}{\micro m}. (\textbf{C}) Longitudinal resistivity calculated for a reference and a \SI{135}{GHz} cavity-embedded Hall bar, as a function of magnetic field. Around filling factor $6+1/2$ we observe a ratio of about~28.}
    \label{fig:D200923-results}
\end{figure}

This sample was processed on a different heterostructure and employing a different design of the resonator with respect to the one presented in the main text. The heterostructure contains a \SI{27}{nm}-wide square quantum well, with a \SI{80}{nm} spacer from  the doping layer, and the 2DES has density \SI{3.2e11}{cm^{-2}} and mobility \SI{2.2e7}{cm^2V^{-1}s^{-1}} (measured at \SI{1.3}{K} without illumination). The cavity is a complementary split-ring resonator (CSRR)~\cite{scalari2012ultrastrong}, and the main difference with the slot antenna resonator comes from the longer length of the current path around the capacitor gap (provided by the arms' shape), which makes the inductance larger. We have fabricated two CSRRs, with resonance frequencies \SI{115}{GHz} and \SI{135}{GHz}, and normalized coupling 45\% and 39\%, respectively. Since the resonant frequency is related to the length of the resonator, we have scaled the distance between the probes accordingly, and fabricated also two different reference Hall bars, with the same voltage probe distance as the respective resonator. The Hall bar width is \SI{50}{\micro m} (see Fig.~\ref{fig:D200923-results}B for the microscope pictures with annotated dimensions). We report in Figs.~\ref{fig:D200923-results}A,C the longitudinal resistivities for both pairs of reference and cavity-embedded Hall bars, and one can notice the cavity-induced suppression of the resistivity peaks at filling factor $6+1/2$ being more than a factor of 20.

\subsubsection{Sample processed on D151202B (D151202B)}\label{subsubsec:D151202B}
Also this sample was processed on another different heterostructure, which contains a \SI{30}{nm}-wide square quantum well, with a \SI{100}{nm} spacer from  the doping layer, and the 2DES has density \SI{2.1e11}{cm^{-2}} and mobility \SI{1.7e7}{cm^2V^{-1}s^{-1}} (measured at \SI{1.3}{K} without illumination). The cavity design is again a CSRR, with \SI{180}{GHz} resonance frequency of the fundamental mode. On this sample we also fabricated a cavity having serrated edges, to investigate the role of edge flatness, as discussed in Section~\ref{sec:cavity_edges}. As already discussed therein, we observe a similar resistivity suppression in the HB embedded in the cavity having flat edges, which is in contrast absent in the sample within the cavity having serrated edges.

\begin{figure}[!h]
    \centering
    \includegraphics[width=\linewidth]{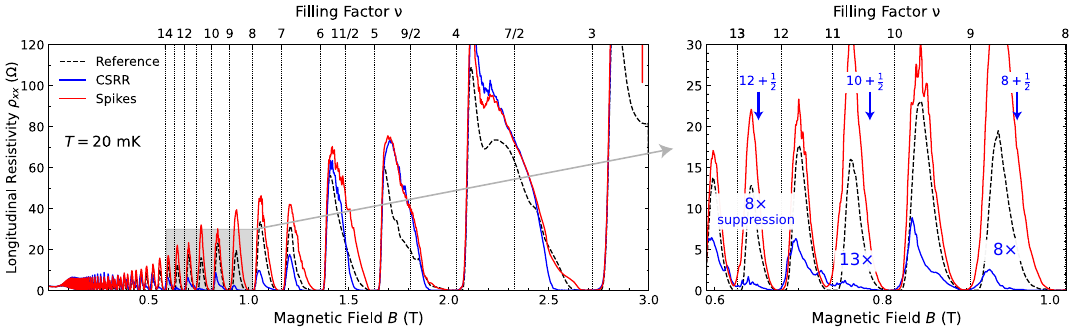}
    \caption{Sample \textbf{D151202B}---Longitudinal resistivity as a function of magnetic field, for the reference HB (black dashed line), the complementary split-ring resonator cavity-embedded HB ("CSRR", blue solid line), and the variation of the latter having serrated edges ("Spikes", red solid line)---see the microscope picture of  Fig.~\ref{fig:role_edges}, bottom panel. The right panel is an enlargement of the grey rectangle highlighted on the left panel. We observe that while the resistivity in the flat edges cavity-embedded sample is about an order of magnitude lower than the reference one, in the serrated edge cavity the effect is absent.}
    \label{fig:spike_cavity}
\end{figure}

\subsubsection{Sample processed on D151202B, investigated in Ref.~\texorpdfstring{\cite{enkner2024enhanced}}{}}
\begin{figure}[!h]
    \centering
    \includegraphics[width=\linewidth]{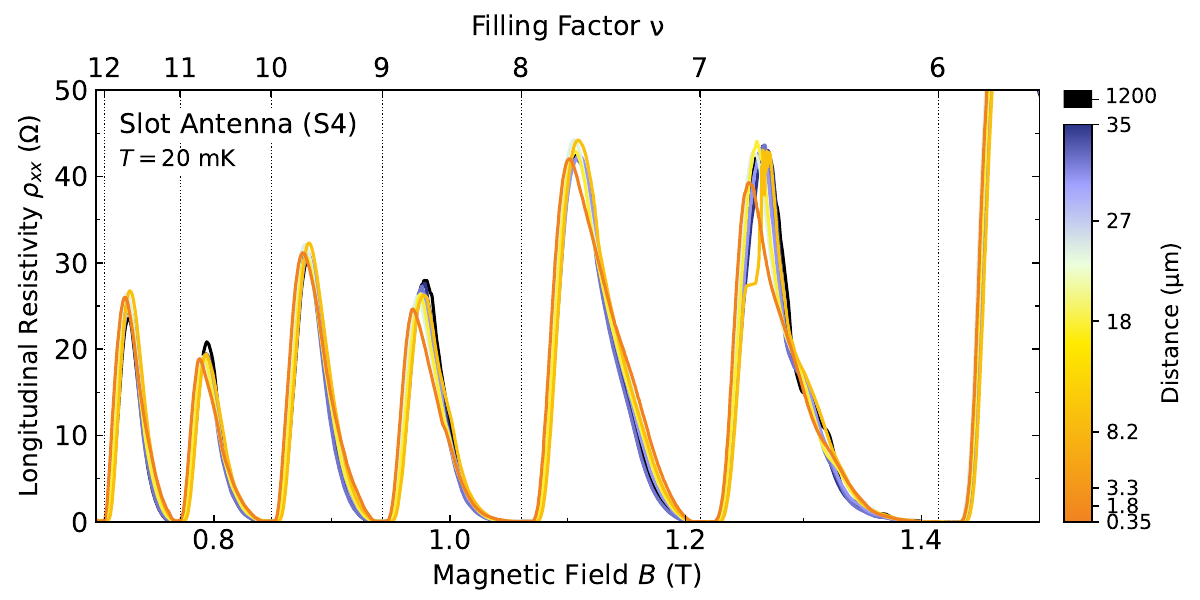}
    \includegraphics[width=\linewidth]{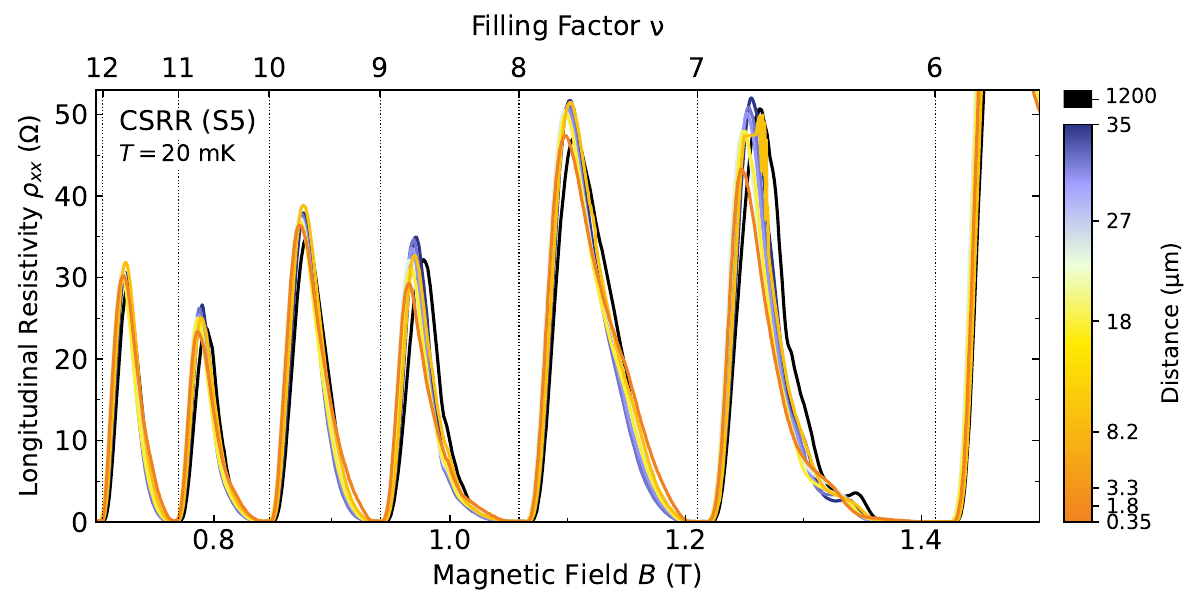}
    \caption{Sample of \textbf{Ref.~\cite{enkner2024enhanced}}---Longitudinal resistivity as a function of magnetic field measured in the hovering cavity experiment reported in Ref.~\cite{enkner2024enhanced}. Curves with different colors refer to the different distances between the hovering cavity and the 2DES, according to the color bar on the right. The top panel refers to the slot antenna cavity, while the bottom one to the complementary split-ring resonator.}
    \label{fig:tipexp}
\end{figure}

The heterostructure from which this sample was fabricated is the same one of Section~\ref{subsubsec:D151202B}. The details on the cavity design and measurement technique are reported in detail in Ref.~\cite{enkner2024enhanced}. Here we limit ourselves to report in Figure~\ref{fig:tipexp} the longitudinal resistivity as a function of magnetic field at values relevant for the observation of the stripe-ordered phase. We indeed see---to a small extent---that reducing the distance between the hovering cavity and the 2DES, i.e.\ increasing the light-matter coupling, the resistivity peaks at half-integer filling factors reduce in amplitude. We remark that the lowest achievable distance was \SI{0.35}{\micro m}, at variance with the cavity-embedding of the Hall bar utilized in the present work.

\clearpage

\section{Temperature study of the odd-integer filling factor minima}
In this section we discuss the thermally activated behaviour of resistivity minima at odd-integer filling factor, which allows us to estimate the Zeeman energy gap, and to show that in the cavity sample this is strongly reduced with respect to the reference one. This confirms our previous results reported in Ref.~\cite{enkner2024enhanced}, and we also comment on the connection with the results of Ref.~\cite{appugliese2022breakdown}. The wealth of effects which we observe in cavity-embedded samples at different filling factor regimes requires indeed that our results are consistent with each other.

In Ref.~\cite{appugliese2022breakdown} we experimentally demonstrated in a Hall bar embedded in a similar subwavelength cavity how the nonlocal nature of vacuum fields could provide, via a vacuum-induced scattering mechanism~\cite{ciuti2021cavity}, a finite resistivity to the otherwise zero-resistance states at high integer filling factors, affecting in particular the odd integer ones. The latter refer indeed to the case in which the Fermi energy lies within Zeeman split states, whose energy gap $\Delta E_\mathrm{Zeeman}$ is much smaller than the cyclotron gap, separating instead occupied and empty states at even integer filling factor regimes. 

In Ref.~\cite{enkner2024enhanced} we further understood that the vacuum-induced scattering---within a single-particle description---induces an effective cavity-mediated electron-electron interaction---at the many-body level--- which affects the magnitude of $\Delta E_\mathrm{Zeeman}$ and strongly reduces it in the odd integer filling factor regime (a behaviour already observed in Ref.~\cite{appugliese2022breakdown}). We can express the energy gap as a function of magnetic field $B$ as 
\begin{equation}\label{eq:ZeemanGap}
\Delta E_\mathrm{Zeeman} = g \mu_B B - \varGamma,
\end{equation}
where $g$ is the effective electronic $g$-factor, which is renormalized by electron-electron interactions~\cite{nicholas1988exchange}, $\mu_B$ is Bohr's magneton, and $\varGamma$ is a parameter that takes into account the Landau level broadening due to disorder~\cite{matthews2005temperature}. Studying the temperature dependence of the longitudinal resistivity minima $\rho_{xx}^\nu$ at the odd integer filling factors we can obtain an estimate of $\Delta E_\mathrm{Zeeman}$: indeed, $\rho_{xx}^\nu$ are thermally activated following 
\begin{equation}\label{eq:arrhenius}
\rho_{xx}^\nu(T) = \rho_{xx}^\nu(0) e^{-T_\text{act} / 2 T} = \rho_{xx}^\nu(0) e^{-E_\text{act} / 2 k_B T},
\end{equation}
where $k_B$ is Boltzmann's constant, $T$ is the temperature, and the activation energy, equivalent to the mobility gap, $E_\mathrm{act} = k_B T_\text{act} = \Delta E_\mathrm{Zeeman}$. The reason for the factor of $1/2$ comes from the law of mass-action, prescribing that the number of electrons excited to the empty Zeeman split state above the Fermi energy is equal to the number of holes left behind in the Zeeman split state below it.

\begin{figure}[!h]
    \centering
    \includegraphics[width=\linewidth]{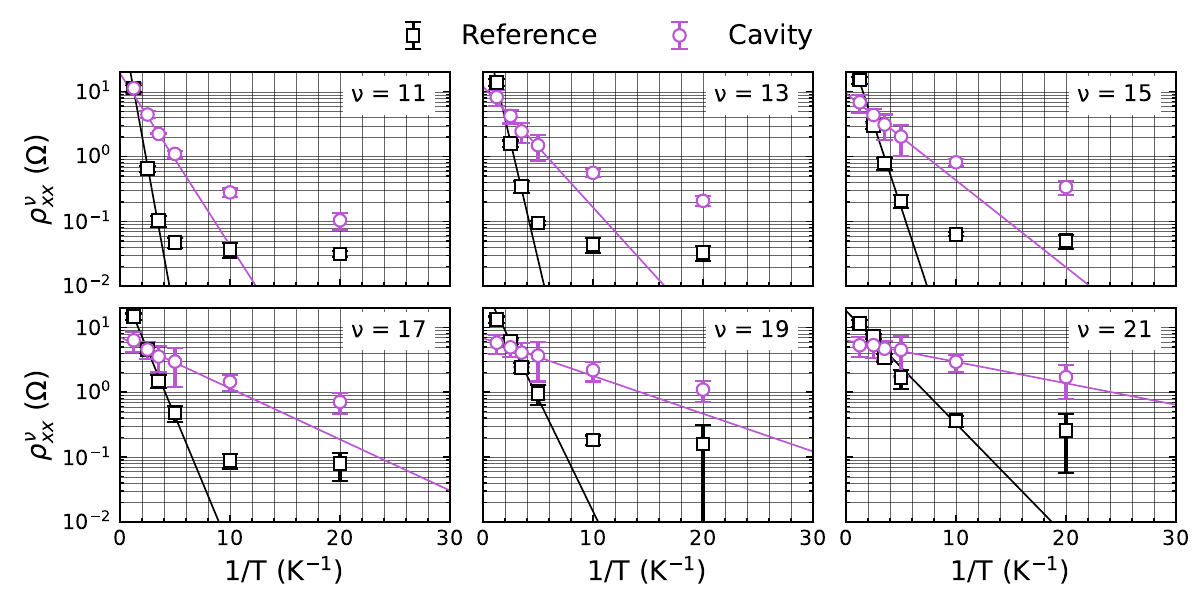}
    \caption{Longitudinal resistivity minima at odd-integer filling factors between 11 and 21 as a function of inverse temperature, for both the reference (empty square markers) and cavity (empty round markers) samples. For temperatures above \SI{0.1}{K} the resistivity minima display an activated behaviour following Eq.~\ref{eq:arrhenius}, which is fitted via the solid black and purple lines, for the reference and cavity sample respectively. The data are obtained as an average between both positive and negative magnetic field (after putting together measurements obtained on the two sides of the Hall bar, as explained in Sec.~\ref{sec:density_inhom}) and both up and down field sweep directions, and the error bars represent their standard deviation.}
    \label{fig:oddminima_activation}
\end{figure}

\begin{figure}[!h]
    \centering
    \includegraphics[width=0.6\linewidth]{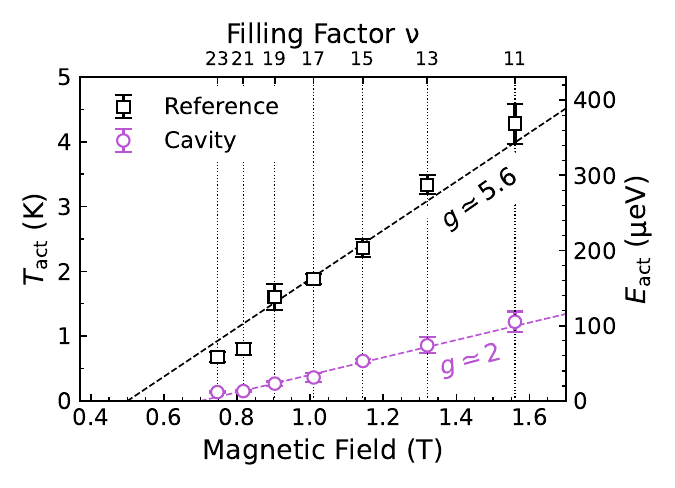}
    \caption{Activation temperature (energy on the right axis) as a function of magnetic field at odd integer filling factors for both reference (black empty squares) and cavity sample (purple empty circles). Notice how the slope, which is proportional to the $g$-factor, is vastly suppressed in the cavity, linking the present work to the findings of Ref.~\cite{enkner2024enhanced}.}
    \label{fig:gfactor}
\end{figure}

In Fig.~\ref{fig:oddminima_activation} we report the longitudinal resistivity minima at odd-integer filling factors between 11 and 21 as a function of inverse temperature, for both reference and cavity samples. The minima follow a temperature activated behaviour following Eq.~\ref{eq:arrhenius} for temperatures above \SI{0.1}{K}, and by a fitting procedure we obtain the activation energy (temperature) which we report in Fig.~\ref{fig:gfactor}, as a function of magnetic field. We observe how the activation energy for the cavity-embedded sample is consistently lower than the one of the reference sample. Moreover, while at low temperatures the resistivity minima for the cavity sample are higher than the one of the reference (as it was demonstrated in Ref.~\cite{appugliese2022breakdown}), increasing the temperature we observe the opposite. By fitting the activation energy to Eq.~\ref{eq:ZeemanGap}, we can extract the value of the $g$-factor, which is strongly suppressed in the cavity with respect to the value in the reference. This result is consistent with what was reported in Ref.~\cite{enkner2024enhanced}, the absolute values being lower due to the higher electron density of the 2DES~\cite{janak1969g}.

\clearpage

A point deserving further investigation is the relationship between the suppression of the $g$-factor and the different behavior of the reference-to-cavity resistivity ratio observed at half-integer filling factors $\nu=2N+1/2$ and $\nu=(2N+1)+1/2$, which we connected in the main text to the different occupation of the spin-split Landau levels (see Fig.~1E in the main text). Specifically, in samples having lower electron density we have observed a lower extent of the $g$-factor suppression at odd integer filling factors, connected with a less pronounced suppression of the resistivity at half-integer filling factors, which could indicate a relationship between the two, not currently taken into account in the present theoretical interpretation.

As a technical note, the temperature was changed by passing current in a \SI{120}{\ohm} resistor attached to the mK plate of the dilution refrigerator, and controlling the measured temperature via a PID feedback loop, allowing us to change it between \SI{20}{} and \SI{800}{mK}. For temperatures below \SI{50}{mK} the reading at finite magnetic field is not reliable, due to the magnetoresistance of the temperature sensor.

\clearpage

\section{Theoretical Interpretation}

Our goal in this section will be to calculate the free energy of the system for different orientations of quantum Hall stripes in the presence of the vacuum fluctuations of the slot antenna cavity. This will permit us to compute the anisotropy \textit{per electron} induced by the interplay of electromagnetic vacuum fluctuations and the quantum Hall stripes, helping us determine how the stripes should align. 

\subsection{Generic Formulation}
We begin by establishing a general formalism for calculating the perturbative contribution to the  free energy arising from the coupling of electronic degrees of freedom and electrodynamic fluctuations of the slot antenna cavity, working within the imaginary time Matsubara formalism~\cite{altland2023condensed}. We consider a generic two-dimensional electron system (2DES) characterized by a current density $\mathbf{J}(\mathbf{r},\tau)$, coupled to the THz cavity induced electromagnetic field described by the vector potential $\textbf{A}(\textbf{r},t)$.  The interaction between this current and the electromagnetic field can be modeled using the  Matsubara action:
\begin{equation}
\mathcal{S}_{\rm int} = \int_0^\beta d\tau \int d^2 \textbf{r}\,  \mathbf{J}(\mathbf{r},\tau) \cdot \textbf{A}(\mathbf{r},\tau),
\label{eqn:action}
\end{equation}
where $\textbf{r} = (x,y)$ and the integration over the imaginary time $\tau$ runs from zero to $\beta = 1/k_B T$. We work within the Weyl Gauge where we set the scalar potential $\phi(\textbf{r},t)=0$. As the vector potential and current fluctuate around zero, we proceed by evaluating the free energy contribution arising from Eq.~\ref{eqn:action}  within  second-order perturbation theory. We present our results for the specific case where the electronic conductivity in the 2DES is local in space, finding the free energy\cite{Chattopadhyay2025}:
\begin{equation}
\label{eqn:cavFE}
F_{\rm anis} = T \sum_{\substack{\omega_m > 0 \\  a \in\{x,y\}}}\omega_m \sigma_{aa}(i \omega_m)  \overline{\langle \textbf{A}^a(\omega_m) \textbf{A}^a(-\omega_m) \rangle}, 
\end{equation}
where the dynamical conductivity tensor, continued to imaginary Matsubara frequencies ($\omega_m  = 2 \pi m/\beta$),  is given by $\sigma_{ij}(i \omega_m)$ and where the sample-averaged Matsubara frequency correlation function of the vector potential arising from the metamaterial cavity is given by  $\overline{\langle \textbf{A}^a(i\omega_m) \textbf{A}^b(-i\omega_m) \rangle} =  \int d^2 \textbf{R} \langle \textbf{A}^a(\mathbf{R},i\omega_m) \textbf{A}^b(\mathbf{R},-i\omega_m) \rangle$.

\subsection{Stripe Anisotropy}

Having articulated how to calculate the free energy arising due to light-matter coupling for a generic 2DES characterized by a local conductivity, we now turn towards computing the free energy for stripe configurations specifically. In particular, let us describe a particular stripe orientation by the angle $\theta$ formed by the wavevector $\hat{\vb{Q}}_{\rm stripe}$ of the density modulation and the $\hat{\vb{y}}$-axis defined in Fig.~1C in the main text (i.e.\ along the short axis of the cavity). For clarity, $\theta = 0$ ($\hat{\vb{Q}}_{\rm stripe} \parallel \hat{\vb{y}}$) implies that the \textit{easy axis} of transport is along the $\hat{\vb{x}}$-axis (hard axis of transport along the $\hat{\vb{y}}$-axis); $\theta = \pi/2$ ($\hat{\vb{Q}}_{\rm stripe} \parallel \hat{\vb{x}}$) implies the reverse.  Experimental findings are consistent with $\theta = 0$.

We now specify the dynamical conductivity of the stripes, generalizing the semi-classical Drude-like transport theory for the dc conductivity for stripes \cite{MacDonald.2000}---a theory that quantifies precisely the intuition of an ``easy" and a ``hard" direction---to finite (imaginary) frequencies: $\sigma_{ij}(i\omega_m) = \frac{1}{1+|\omega_m|\tau} \sigma_{ij}(\omega_m=0)$, where $\sigma_{ii}(0) = \frac{e^2 \sqrt{R}}{2 h}$ and $\sigma_{jj}(0) = \frac{e^2}{2 h  \sqrt{R}}$;  where the ``easy" $i$ axis is given by $\mathbf{\hat{e}}_i = \cos(\theta) \mathbf{\hat{e}}_x + \sin(\theta) \mathbf{\hat{e}}_y$  and the ``hard" axis $j \perp i$; where the anisotropy ratio is $R = \frac{\sigma_{ii}}{\sigma_{jj}} \gg 1$. Here the relevant transport scattering time is  $\tau = \frac{a \sqrt{R}}{v_F}$\cite{MacDonald.2000}, where $v_F$ is the Fermi velocity and $a$ is the inter-stripe separation (i.e., $|\hat{\vb{Q}}_{\rm stripe}| = \frac{2 \pi}{a})$. We note that the Drude transport within this model arises from disorder that gives rise to inter-stripe scattering, introducing both a finite conductivity for transport along the hard direction and a finite resistance to transport along the easy direction. It is precisely this disorder that allows $q=0$ components of the electric field to couple to the conductivity of the electrons, eliding the restrictions imposed by Kohn's theorem. Leveraging this model for the dynamical conductivity---in the limit of $R \gg 1$---the stripe anisotropy per particle in the highest partially filled Landau level arising from vacuum fluctuations of the THz cavity is given by: 
\begin{equation}
\label{eqn:cavFE_delta}
\Delta F_{\rm anis} = \frac{T e^2 \sqrt{R}}{2 h} \times  \sum_{\substack{\omega_m>0}}\frac{\omega_m}{1+\omega_m \tau}  (\overline{\langle \textbf{A}^y(i\omega_m) \textbf{A}^y(-i\omega_m) \rangle}-\overline{\langle \textbf{A}^x(i\omega_m) \textbf{A}^x(-i\omega_m) \rangle}) \cos(2 \theta), 
\end{equation}

With a description of the electronic degrees of freedom in place, we articulate a model that captures the vacuum fluctuations of the slot antenna cavity, aiming to characterize $\overline{\langle \textbf{A}^a(i\omega_m) \textbf{A}^a(-i\omega_m) \rangle}$ through a decomposition in terms of the modes of the cavity.  Neglecting dissipation, $\overline{\langle \textbf{A}^a(i\omega_m) \textbf{A}^a(-i\omega_m)} = \frac{1}{\epsilon_0 \epsilon_R d} \sum_{\lambda} \frac{f_{\lambda}^a}{\omega_m^2+\omega_\lambda^2}$, where $\lambda$ runs over the modes of the cavity; where $f_{\lambda}^ a$ quantifies how much of the planar component of the mode $\lambda$ is polarized along direction $a$ (e.g. $f_{\lambda}^x = \frac{\int d^2 R |A_\lambda^x(R)|^2}{\int d^2 R |A_\lambda^x(R)|^2+|A_\lambda^y(R)|^2}$); where $\epsilon_R = 13.1$ is the dielectric constant of  \ce{GaAs}; where the effective sub-wavelength confinement along the $z$-axis is codified through $d \approx 100 \textrm{nm}$.  An epistemically modest estimate of the anisotropy arises from considering  the contributions from just the fundamental mode of the resonator, given that this is the only mode that has been unequivocally demonstrated to have strong coupling to electrons in the 2DES~\cite{scalari2012ultrastrong}. Here, as the mode is entirely polarized along the $\hat{\vb{y}}$-axis ($f^{y}_{\rm fund} = 1, f^x_{\rm fund}=0$) it is clear that stripes will align with angle $\theta = 0$. Estimating this anisotropy leads to $\Delta F_{\rm anis} \approx 100 $ K $\approx 9$ meV.

\bibliography{bibliography,refs-curtis}